\documentclass[preprint,preprintnumbers,amsmath,amssymb]{revtex4-1}
\usepackage{graphicx}% Include figure files
\usepackage{dcolumn}% Align table columns on decimal point
\usepackage{bm}% bold math
\usepackage{amssymb,amsfonts,amsmath}
\usepackage{graphicx}
\usepackage{threeparttable}
\usepackage{float}

\begin{document}
%\linenumbers
%\setpagewiselinenumbers
%\modulolinenumbers[5]
%\linenumbers

%\preprint{APS/123-QED}

\title{\bf{Structural and Magnetic Reconstruction of Thermodynamically Stable CaMnO$_{3}$ (001) Surfaces } \\[11pt] }

\author{Diomedes Saldana-Greco, Chan-Woo Lee, Doyle Yuan, and Andrew M. Rappe}

\affiliation{
  The Makineni Theoretical Laboratories, Department of Chemistry, 
  University of Pennsylvania, Philadelphia, PA 19104-6323, USA\\ }

\date{\today}% It is always \today, today,
             %  but any date may be explicitly specified

                             % Classification Scheme.
%\keywords{Suggested keywords}%Use showkeys class option if keyword

\begin{abstract}

The relative thermodynamic stability of surface reconstructions 
including vacancies, adatoms and additional layers on both 
CaO and MnO$_{2}$ terminations is calculated to predict the 
surface phase diagram of ($\sqrt{2}\times\sqrt{2}$)\emph{R}45$^{\circ}$ 
CaMnO$_{3}$ (001) using \emph{ab initio} thermodynamics.
Stoichiometric and nonstoichiometric reconstructions are considered.
A set of boundary conditions driven by the binary and ternary sub-phases 
from CaMnO$_{3}$ defines its bulk stability region, enclosing the stable surface 
reconstructions that can be in equilibrium with the bulk.
Most of the surfaces take the magnetic ordering of the bulk ground state, G-type 
antiferromagnetic (AFM); however, some of the MnO$_{2}$-terminated surfaces are more stable 
when the surface layer spins flip.
At 573 K, the stoichiometric CaO- and MnO$_{2}$-terminated surfaces are predicted to be 
thermodynamically stable, as well as a CaO-terminated surface reconstruction where half the 
Ca are replaced by Mn. 
The MnO$_{2}$-terminated surface reconstructions dominate the phase diagrams 
at high temperatures, including phases with MnO and MnO$_{2}$ adatoms 
per surface unit cell.  
 
\end{abstract}

                              %display desired
\maketitle

\section*{\label{sec:level1} I. Introduction}

Perovskite CaMnO$_{3}$ has recently been the subject of increasing 
interest due to its strain-assisted multiferroicity~\cite{Bhattacharjee09p117602, Wang12p17005}, 
thermoelectric efficiency~\cite{Urata07p535, Koumoto10p363}, 
colossal magnetoresistance~\cite{Zeng99p8784}, and catalytic 
properties~\cite{Kawashima08p3439}.
Its high catalytic activity for oxidation reactions and structural
similarity to important biological catalysts has motivated studies 
on CaMnO$_{3}$ as a potential catalyst for water oxidation~\cite{Najafpour12p1}.    
While atomic insight into the CaMnO$_{3}$ surface is imperative if 
its potential application as a catalyst is to be realized, very little 
has been investigated in this direction.
Most of the previous studies have focused on magnetic 
properties~\cite{Filippetti99p4184, Filippetti00p11571, Nguyen11p3613} 
of the stoichiometric cubic CaMnO$_{3}$ surface. 
In this study, we determine the atomic and magnetic low-energy surface structures 
of orthorhombic \emph{Pnma} CaMnO$_{3}$ (001) as functions of environmental variables, 
developing surface phase diagrams by combining density functional theory (DFT) 
and thermodynamics~\cite{Reuter01p035406}. 
Various surface reconstructions for both CaO- and MnO$_{2}$-terminated surfaces are explored within 
a  ($\sqrt{2}\times\sqrt{2}$)\emph{R}45$^{\circ}$ surface periodicity. 
For structures with Mn atoms at the surface, we considered multiple magnetic spin configurations to 
ensure that the surface stability accounts for the potential magnetic reconstruction at the surface. 
The surface free energies are computed to generate the surface phase diagram within the bulk 
stability range. 
To illustrate the surface stability at a range of temperatures, the vibrational free energy 
is included in the calculation of the bulk stability region. 
Through our surface examination, we predict thermodynamically stable surface reconstructions 
at a range of conditions.   

\section*{\label{sec:level1} II. Methodology}

\subsection*{\label{sec:level2} A. Computational Methods}

DFT calculations were performed using the PBEsol~\cite{Perdew08p136406} 
generalized gradient approximation (GGA), as implemented in the 
{\sc Quantum Espresso}~\cite{Giannozzi09p395502} computer code.
Perdew and collaborators have shown that the PBEsol functional describes 
solids and surfaces accurately~\cite{Perdew08p136406}.
PBEsol+\emph{U} was also used to test electronic properties, with 
\emph{U}$_{\textrm{eff}}$ = 7.23 eV, in line with previous 
work~\cite{Luo07p036402} and determined with the linear response scheme~\cite{Cococcioni05p035105}.
However, adding strong correlations to CaMnO$_{3}$ provides unsatisfactory 
description of the magnetic properties, as has been established by 
Luo \emph{et al.}~\cite{Luo07p036402}.  
Spin-polarized electronic densities were used in all 
calculations, treating the magnetic moments as collinear in all cases, with additional 
non-collinear (NC) calculations for some Mn-terminated surfaces and for the ternary sub-phases. 
All atoms are represented by norm-conserving, optimized~\cite{Rappe90p1227}, 
designed nonlocal~\cite{Ramer99p12471} pseudopotentials with spin-orbit 
interaction (for NC magnetism).
The pseudopotentials are generated with the {\sc opium} package~\cite{Opium}, 
treating the 3\emph{s}, 3\emph{p}, 3\emph{d}, 4\emph{s}, and 4\emph{p} of Ca, the 3\emph{s}, 
3\emph{p}, 3\emph{d}, 4\emph{s}, and 4\emph{p} of Mn, and the 2\emph{s} and 
2\emph{p} of O as valence states. 
An accurate description of the magnetic properties of all the systems studied 
is achieved when nonlinear core-valence interaction is included in the Mn 
pseudopotential by the partial core correction scheme~\cite{Fuchs99p67, Porezag99p14132}.  
The CaMnO$_{3}$ magnetic ground state is collinear G-type antiferromagnetic (AFM) with 
an observed N\'{e}el temperature of about 130 K~\cite{Poeppelmeier82p71}, estimating 
the magnetic interaction energy to be around 6.6 meV~\cite{Bhattacharjee08p255229}.
Therefore, all calculations are run with a 70 Ry plane-wave energy cutoff to 
ensure accuracy for the small relative energies among the different magnetic 
configurations.
The Brillouin zone is sampled using a $4\times4\times1$ Monkhorst-Pack~\cite{Monkhorst76p5188} 
$k$-point mesh for surface structures based on energy convergence for bulk 
CaMnO$_{3}$.
For binary and ternary sub-phases, a dense enough set of $k$-points was 
used so that the total energy is converged.
The ionic relaxation parameters are chosen so that the forces in the surface 
and bulk structures are lower than 2 meV/{\AA}. 
We use density functional perturbation theory~\cite{Baroni01p515, Gonze95p1096} 
to calculate phonon frequencies and vibrational displacement 
vectors at the $\Gamma$ point for all binary and ternary sub-phases in order 
to include the vibrational free energy contribution into the definition of the 
free energy.
Translational symmetry is ensured by enforcing the three translational acoustic sum rules 
requiring that the acoustic modes have zero frequency. 
This is accomplished through small corrections to the diagonal elements 
of the dynamical matrix. 

\subsection*{\label{sec:level2} B. CaMnO$_{3}$ (001) Surface Structures}

\begin{figure}
\centering
\includegraphics[width=0.75\textwidth, angle =90]{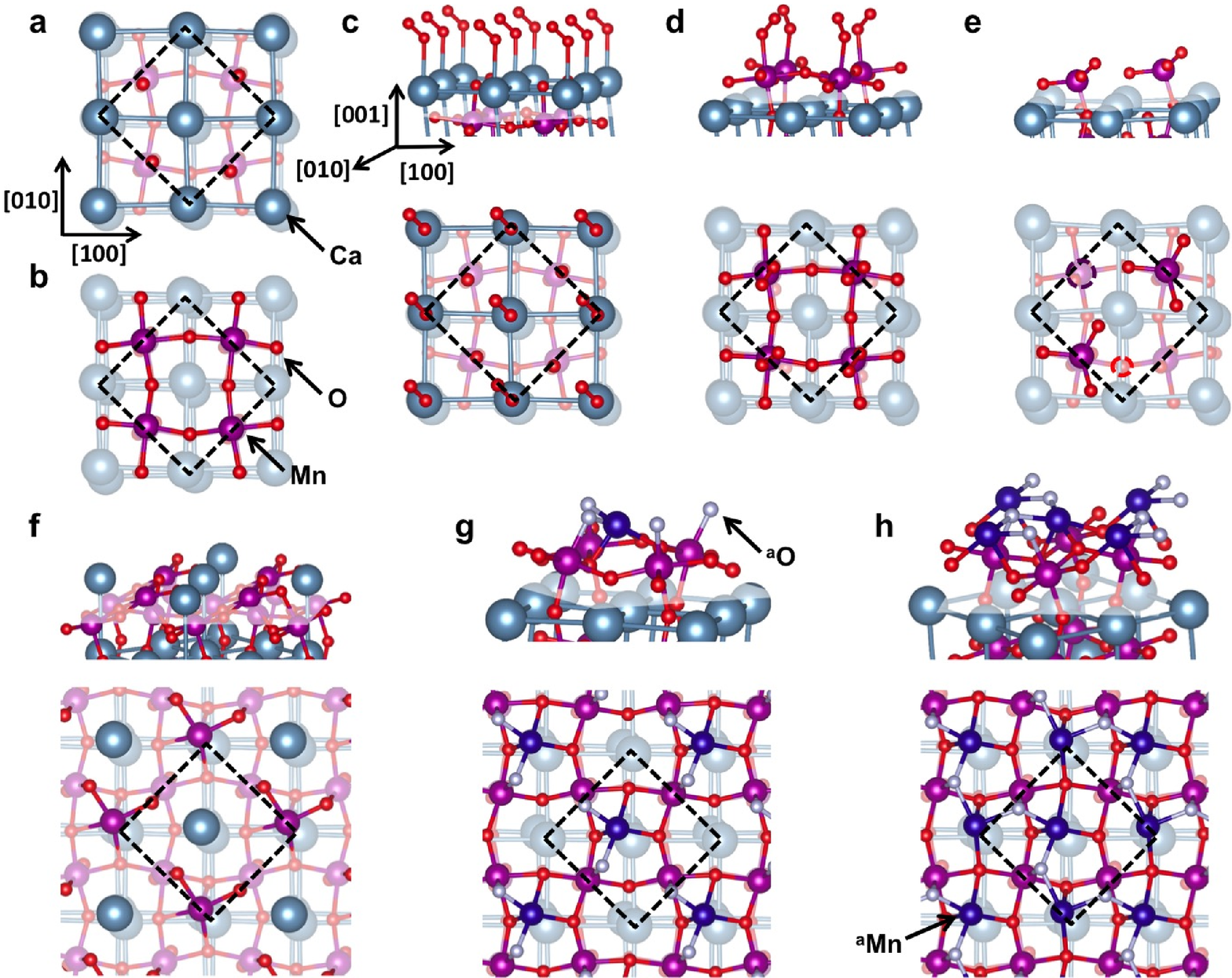}
\caption{Relaxed atomic structures of CaMnO$_{3}$ (001) thermodynamically stable surface 
reconstructions: Stoichiometric (a) CaO-terminated,  (b) MnO$_{2}$-terminated, (c) CaO+2.0O, 
(d) MnO$_{2}$+2.0O, (e) MnO$_{2}$-0.5Mn-0.5O, (f) CaO-0.5Ca+0.5Mn, 
(g) MnO$_{2}$+0.5Mn+1.0O, and (h) MnO$_{2}$+1.0Mn+1.0O. Large light blue spheres: Ca; 
medium purple spheres: Mn; and small red spheres: O. All surface reconstructions are within 
a ($\sqrt{2}\times\sqrt{2}$)\emph{R}45$^{\circ}$ surface periodicity, denoted with black dashed 
lines. Top and side views of the surface are shown for reconstructions on figures (c-h). For (e), 
the dashed purple and red circles represent the Mn vacancy and O vacancy, respectively. 
For (g,h), the $^{a}$O and $^{a}$Mn atoms are O and Mn adatoms colored white and blue, 
respectively.} 
\label{Fig.1}
\end{figure}

The (001) surface of CaMnO$_{3}$ is the lowest energy, since it 
consists of alternating CaO and MnO$_{2}$ planes, each of which is neutral 
in an ionic picture~\cite{Tasker79p4977}.
We construct surface structures with different compositions 
and reconstructions for both CaO and MnO$_{2}$ terminations, 
with symmetrical slabs of 7-9 layers and a vacuum gap between slabs  
of $\approx$15 {\AA}.
All atoms are fully relaxed. 
Over 80 surface terminations are made by varying the stoichiometry of 
Ca, Mn and O including additional Ca$_{x}$O$_{y}$ and Mn$_{x}$O$_{y}$ 
layers in the ($\sqrt{2}\times\sqrt{2}$)\emph{R}45$^{\circ}$ surface 
symmetry. 
The positions of adatoms and additional layers are strategically 
selected; Ca and Mn atoms are placed on top of O atoms, and O 
species on top of Ca and Mn atoms. 
Each surface termination is identified based on its stoichiometric 
termination (either CaO or MnO$_{2}$) and whether its composition 
involves vacancies (-) or adatoms (+).
As an illustration, CaO+2.0O refers to a CaO-terminated surface with 
two oxygen adatoms per primitive unit cell of CaO. 
The number of adatoms and/or vacancies is tabulated per formula unit of 
either CaO or MnO$_{2}$, even though there are two formula units in the surface 
super cell. 
Similarly, $\mbox{MnO$_{2}$-0.5Mn-0.5O}$ refers to a MnO$_{2}$-terminated surface 
with one Mn vacancy and one oxygen vacancy per two MnO$_{2}$ units.
Illustrations of the thermodynamically stable surface terminations after relaxation 
are shown in Fig.\ \ref{Fig.1}.
These surface phases include (a,b) the stoichiometric surfaces, (c,d) their full 
surface coverage with adsorbed oxygen molecules, (e) the MnO$_{2}$-terminated surface 
with one Mn vacancy and one oxygen vacancy, as well as (f) a CaO-terminated surface 
where a Mn adatom occupies a Ca vacancy.
The last two surface phases (g,h) are stable at high temperatures but represent 
very exotic surface structures, as the $^{a}$Mn adatoms and $^{a}$O adatoms have 
different coordination than the stoichiometric atoms, potentially indicating active sites.   

\begingroup
\begin{table}
\caption{Energies per formula unit and magnetic moments  
on Mn atoms for bulk CaMnO$_{3}$ in different magnetic phases.}
\begin{ruledtabular}
\centering
\begin{tabular}{lccccc}%
                                  &   FM   &   A-AFM   &   C-AFM   &   G-AFM   \\ \hline
Energy (meV/formula unit)         &   38   &     12    &     4     &    0      \\
Magnetic moment ($\mu$$_{B}$/Mn)  &  2.75  &    2.68   &   2.64    &   2.61    \\

\end{tabular}
\end{ruledtabular}
\end{table}
\endgroup

The ferromagnetic (FM) and three AFM phases of bulk CaMnO$_{3}$ were analyzed.
The A-AFM state has FM planes of alternating direction along the \emph{z}-axis. 
The C-AFM arrangement consists of a Mn with nearest neighbors 
of opposite spin in the plane, and parallel spins between one plane and 
the next, while G-type AFM arranges the spins so that 
all nearest neighbors have opposite spin orientation.
The magnetic ground state of bulk CaMnO$_{3}$ is found to be G-AFM 
\emph{Pnma} insulator, as reported in previous theoretical and experimental 
studies~\cite{Pickett96p1146, Bhattacharjee08p255229, Poeppelmeier82p71}.
Our computed electronic band gap is 0.65 eV, significantly underestimating the 
experimental value of 3.1 eV~\cite{Jung97p15489} but in close agreement 
with previous computational work~\cite{Tsukahara10p214108,Pickett96p1146}
The atom-projected magnetic moments of Mn atoms in each magnetic phase are shown 
in Table 2 as well as the relative energies with respect to the stable G-AFM phase, 
which are in very good agreement with those of a recent projector augmented wave 
calculation using the PBE functional~\cite{Tsukahara10p214108}.   
The surface calculations are performed with G-AFM ordering, magnetic 
spin flip at the surface and other possible magnetic spin configurations for surfaces 
with complex Mn$_{x}$O$_{y}$ reconstructions (See Supplementary Material).  
The magnetic spin flip configuration is studied for all MnO$_{2}$-based surfaces since 
stoichiometric MnO$_{2}$ is found to be more energetically favorable when all the surface 
spins flip (making the top and second layers FM aligned)~\cite{Filippetti99p4184}.

\subsection*{\label{sec:level2} C. Thermodynamic Stability}

The free energies of CaMnO$_{3}$ (001) surfaces are calculated 
with the thermodynamic approach described in this section.
In this model, the CaMnO$_{3}$ bulk is considered to be in contact 
with an atmosphere composed of all its components in equilibrium. 
This implies that the system acts as a reservoir where removal 
and/or addition of any of its components at the surface leads 
to a range of thermodynamically stable reconstructions and 
compositions at different conditions.    
The oxygen chemical potential, $\mu$$_{\textrm{O}}$, is used to correlate 
the surface stability with environmental conditions such as temperature, 
\emph{T}, and oxygen partial pressure, \emph{p}(O$_{2}$). 
The surface free energy, $\Omega^i$, of an individual surface 
slab, \emph{i}, is defined as the excess amount of free energy 
needed to create the surface from its bulk form~\cite{Reuter01p035406},  
\begin{eqnarray}
\Omega^i &=&\frac{1}{2A}\left[G^i_{\textrm{slab}}-\sum_{j}N^i_{j}\mu_{j}\right] \nonumber \\
         &=&\frac{1}{2A}\left[G^i_{\textrm{slab}}-N^i_{\textrm{Ca}}\mu^i_{\textrm{Ca}}
-N^i_{\textrm{Mn}}\mu^i_{\textrm{Mn}}-N^i_{\textrm{O}}\mu^i_{\textrm{O}}\right].
\end{eqnarray}
The Gibbs free energy of the slab is denoted by G$^i_{\textrm{slab}}$.
The term $\displaystyle\sum\limits_{j}N^{i}_j\mu_{j}$ represents the 
free energy of the material in slab \emph{i} as the sum 
of the chemical potentials from the bulk of each component, $\mu_{j}$, 
multiplied by the number of its atoms in each slab, \emph{N}$^{i}_{j}$~\cite{Cahn73p3}. 
The factor of $\frac{1}{2}$ is included to account for top and bottom 
slabs. 
The surface free energy is reported as energy per unit area.
The $\Gamma^i_{m,n}$ term shown in Eq.\ (2) accounts for the 
off-stoichiometric atoms of component \emph{n} with respect to 
component \emph{m} in any surface slab \emph{i} by relating 
the number of atoms of components \emph{N}$_{m}$ and \emph{N}$_{n}$ in the 
slab to their ratio in the bulk,
\begin{eqnarray}
\Gamma^i_{m,n}=\frac{1}{2A}\left(N^i_{n}-N^i_{m}\frac{N^{\textrm{bulk}}_n}{N^{\textrm{bulk}}_m}\right).
\end{eqnarray}
In our case, the non-stoichiometric surface components were defined 
with respect to Mn atoms.
Defining $\Gamma^i_{m,n}$ with respect to Ca atoms does not change 
the results.
Next, Eqs.\ (1) and (2) are merged to express the Gibbs surface free 
energy as
\begin{eqnarray}
\Omega^i &=&\frac{1}{2A}\left[G^i_{\textrm{slab}}-N^i_{\textrm{Mn}}(\mu_{\textrm{Ca}}+\mu_{\textrm{Mn}}+3\mu_{\textrm{O}})\right] \nonumber \\
         && -\Gamma^i_{\textrm{Mn,Ca}}\mu_{\textrm{Ca}}-\Gamma^i_{\textrm{Mn,O}}\mu_{\textrm{O}}.
\end{eqnarray}

Since the system acts as a thermodynamic reservoir, the chemical
potentials of CaMnO$_{3}$ components are not independent;
they are related to the chemical potential of CaMnO$_{3}$ crystal, 
$\mu$$_{\textrm{CaMnO$_{3}$}}$.
The Gibbs free energy of bulk CaMnO$_{3}$ per formula unit, 
\emph{g}$^{\textrm{bulk}}_{\textrm{CaMnO$_{3}$}}$, is then equal to 
$\mu_{\textrm{CaMnO}_{3}}$ because the slabs must be 
in equilibrium with the bulk as shown below:
\begin{eqnarray}
\mu_{\textrm{Ca}}+\mu_{\textrm{Mn}}+3\mu_{\textrm{O}}=\mu_{\textrm{CaMnO$_{3}$}}=g^{\textrm{bulk}}_{\textrm{CaMnO$_{3}$}}.
\end{eqnarray}
Inserting the relationship above into Eq.\ (3) simplifies the surface 
free energy to
\begin{eqnarray}
\Omega^i &=&\frac{1}{2A}\left[G^i_{\textrm{slab}}-N^i_{\textrm{Mn}}g^{\textrm{bulk}}_{\textrm{CaMnO$_{3}$}}\right] \nonumber \\
         &&  -\Gamma^i_{\textrm{Mn,Ca}}\mu_{\textrm{Ca}}-\Gamma^i_{\textrm{Mn,O}}\mu_{\textrm{O}}.
\end{eqnarray}

The bulk stability region is defined by a set of boundary 
conditions~\cite{Lee11p115418, Heifets11p491, Kolpak08p036102, Levchenko08p256101}.
First, Ca and Mn are not allowed to form metallic precipitates by satisfying the following 
conditions:
\begin{eqnarray} 
\mu_{\textrm{Ca}}                     &\le& g^{\textrm{bulk}}_{\textrm{Ca}}, \\ 
\mu_{\textrm{Mn}}                     &\le& g^{\textrm{bulk}}_{\textrm{Mn}}. 
\end{eqnarray}
Second, the bulk is stable, while binary metal oxides derived 
from its components do not precipitate:
\begin{eqnarray}
\mu_{\textrm{Ca}}+\mu_{\textrm{O}}              &\le& g^{\textrm{bulk}}_{\textrm{CaO}}, \\
\mu_{\textrm{Mn}}+\mu_{\textrm{O}}              &\le& g^{\textrm{bulk}}_{\textrm{MnO}}, \\
3\mu_{\textrm{Mn}}+4\mu_{\textrm{O}}            &\le& g^{\textrm{bulk}}_{\textrm{Mn$_{3}$O$_{4}$}}, \\
2\mu_{\textrm{Mn}}+3\mu_{\textrm{O}}            &\le& g^{\textrm{bulk}}_{\textrm{Mn$_{2}$O$_{3}$}}, \\
\mu_{\textrm{Mn}}+2\mu_{\textrm{O}}             &\le& g^{\textrm{bulk}}_{\textrm{MnO$_{2}$}}. 
\end{eqnarray}
Third, there should be no precipitation of other ternary sub-phases 
(Ca$_{x}$Mn$_{y}$O$_{z}$) from the bulk CaMnO$_{3}$.
The experimental phase diagram of bulk CaMnO$_{3}$ illustrates 
the importance of considering these ternary sub-phases, as they can 
co-exist with CaMnO$_{3}$~\cite{Horowitz78p1359, Balakirev06pS49}.
Consequently, the following conditions must be met:
\begin{eqnarray}
\mu_{\textrm{Ca}}+2\mu_{\textrm{Mn}}+4\mu_{\textrm{O}}   &\le& g^{\textrm{bulk}}_{\textrm{CaMn$_{2}$O$_{4}$}}, \\
\mu_{\textrm{Ca}}+7\mu_{\textrm{Mn}}+12\mu_{\textrm{O}}  &\le& g^{\textrm{bulk}}_{\textrm{CaMn$_{7}$O$_{12}$}}, \\
2\mu_{\textrm{Ca}}+\mu_{\textrm{Mn}}+4\mu_{\textrm{O}}   &\le& g^{\text{bulk}}_{\textrm{Ca$_{2}$MnO$_{4}$}}, \\
3\mu_{\textrm{Ca}}+2\mu_{\textrm{Mn}}+7\mu_{\textrm{O}}  &\le& g^{\textrm{bulk}}_{\textrm{Ca$_{3}$Mn$_{2}$O$_{7}$}}, \\
4\mu_{\textrm{Ca}}+3\mu_{\textrm{Mn}}+10\mu_{\textrm{O}} &\le& g^{\textrm{bulk}}_{\textrm{Ca$_{4}$Mn$_{3}$O$_{10}$}}.
\end{eqnarray}
The DFT energy is taken as the Helmholtz free energy at 0 K with zero-point 
energy, so the Gibbs free energy and the DFT energy are 
related as follows~\cite{Reuter01p035406},
\begin{eqnarray}
G=E^{\textrm{total}}+F^{\textrm{vib}}-TS^{\textrm{conf}}+pV\approx E+F^{\textrm{vib}}.
\end{eqnarray}
The energetic contributions provided by the \emph{pV} and the configurational 
entropy \emph{TS}$^{\textrm{conf}}$ terms to the thermodynamic stability 
are considered to be negligible, as has been proven in previous 
studies~\cite{Lee11p115418, Heifets11p491, Kolpak08p036102, Levchenko08p256101}.  
The term \emph{F}$^{\textrm{vib}}$, which is equal to 
\emph{E}$^{\textrm{vib}}$-\emph{TS}$^{\textrm{vib}}$, is calculated 
from the vibrational modes of the system.
The phonon density of states, $\sigma$($\omega$), is used to express 
\emph{F}$^{\textrm{vib}}$ as an integral over frequency, $\omega$:  
\begin{eqnarray}
F^{\textrm{vib.}}=\int d\omega F^{\textrm{vib.}}(T,\omega)\sigma(\omega).
\end{eqnarray}
The analytical expression of the harmonic quantum vibrational free energy per mode as a 
function of temperature and frequency, \emph{F}$^{\textrm{vib}}$(\emph{T},$\omega$), 
can be written as 
\begin{eqnarray}
F^{\textrm{vib.}}(T,\omega)=\frac{1}{2}\hbar\omega-k_{B}T\ln\left(1-e^{-\frac{\hbar\omega}{k_{B}T}}\right).
\end{eqnarray}
Once the Gibbs free energy of each compound is calculated, the 
stability region of bulk CaMnO$_{3}$ as a function of \emph{T} can be 
evaluated~\cite{Lee11p115418}.
The vibrational free energy is included for all bulk phases while it is neglected 
for all slab structures~\cite{Heifets11p491, Kolpak08p036102, Levchenko08p256101}.
The surface vibrational behavior is complex, and the vibrational modes could 
be affected by the electronic redistribution at the surface region; however, 
it has been shown for other systems that the phonon frequencies do not 
vary the order of stable surface phases~\cite{Lee11p115418}. 
The chemical potential for each component of CaMnO$_{3}$ is defined 
as its difference from the total energy of its reference state:
\begin{eqnarray}
\Delta\mu_{\textrm{Ca}}=\mu_{\textrm{Ca}}-\left(E^{\textrm{bulk}}_{\textrm{Ca}}+F^{\textrm{vib,bulk}}_{\textrm{Ca}}\right), \\
\Delta\mu_{\textrm{Mn}}=\mu_{\textrm{Mn}}-\left(E^{\textrm{bulk}}_{\textrm{Mn}}+F^{\textrm{vib,bulk}}_{\textrm{Mn}}\right), \\              
\Delta\mu_{\textrm{O}}=\mu_{\textrm{O}}-\frac{1}{2}\left(E^{\textrm{gas}}_{\textrm{O$_{2}$}}+F^{\textrm{vib,gas}}_{\textrm{O$_{2}$}}\right).
\end{eqnarray}
For the metals Ca and Mn, the reference energy is the DFT total energy and 
vibrational free energy of the elemental bulk crystal.
For oxygen, the reference is the energy of an O atom 
in an isolated O$_{2}$ molecule.
The \emph{E}$^{\textrm{gas}}_{\textrm{O$_{2}$}}$ is corrected so that 
\emph{E}$^{\textrm{gas}}_{\textrm{O$_{2}$}}$ = 2\emph{E}$_{\textrm{O}}$+
\emph{E}$^{\textrm{expt}}_{\textrm{binding}}$~\cite{Kolpak08p036102} 
since a known shortcoming of DFT is the overestimation of binding 
energies~\cite{Furche01p195120}, particularly for double bonds~\cite{Walter99p11, Grinberg02p2264}.

\begingroup
\begin{table}
\begin{threeparttable}
\caption{Experimental and computed formation energies (eV/formula unit) for
binary and ternary sub-phases defining the bulk stability region of CaMnO$_{3}$.
The calculated formation energies for all Mn-based phases were performed
with their corresponding magnetic ground-state structure, noted next to the compound.
The experimental values are obtained from Ref.\ ~\cite{Fritsch96p1761}.}
\begin{ruledtabular}
\centering
\begin{tabular}{lccc}%
                           &           Experimental   &           Theoretical         \\ \hline
CaO                        &           -6.59          &           -6.71          \\
MnO (AFM-II$^{\emph{a}}$)                        &           -3.99          &           -4.06          \\
Mn$_{3}$O$_{4}$ (FiM-III$^{\emph{b}}$)            &          -14.38          &          -15.46          \\
Mn$_{2}$O$_{3}$ (FM)           &           -9.93          &          -11.36          \\
MnO$_{2}$ (AFM)                 &           -5.39          &           -6.60          \\
CaMn$_{2}$O$_{4}$ (AFM)         &            ---           &          -18.23          \\
CaMn$_{7}$O$_{12}$ (NC-AFM)        &            ---           &          -47.01          \\
Ca$_{2}$MnO$_{4}$ (G-AFM)         &            ---           &          -20.56          \\
Ca$_{3}$Mn$_{2}$O$_{7}$ (G-AFM)   &            ---           &          -34.34          \\
Ca$_{4}$Mn$_{3}$O$_{10}$ (G-AFM)  &            ---           &          -48.12          \\
CaMnO$_{3}$ (G-AFM)               &          -12.83          &          -13.79          \\

\end{tabular}
\end{ruledtabular}
\begin{tablenotes}
\item [\emph{a}]FM layers in the [111] plane and successive antiparallel layers.
\item [\emph{b}]Ferrimagnetic with Mn atoms ($\uparrow$$\uparrow$$\downarrow$$\downarrow$$\uparrow$$\uparrow$) in the corresponding order as Ref.\ ~\cite{Fritsch96p1761}.
\end{tablenotes}
\end{threeparttable}
\end{table}
\endgroup

Combining the chemical potential expressions, Eq.\ (21-23), and the 
approximation to the Gibbs free energy for the slabs and CaMnO$_{3}$ bulk 
leads to the following expression for the Gibbs surface 
free energy,
\begin{eqnarray}
\Omega^i &=& \frac{1}{2A}\left[E^i_{\textrm{slab}}-N^i_{\textrm{Mn}}\left(E^{\textrm{bulk}}_{\textrm{CaMnO$_{3}$}}+F^{\textrm{vib,bulk}}_{\textrm{CaMnO$_{3}$}}\right)\right] \nonumber \\
         &&  -\left[\Gamma^i_{\textrm{Mn,Ca}}\left(\Delta\mu_{\textrm{Ca}}+E^{\textrm{bulk}}_{\textrm{Ca}}+F^{\textrm{vib,bulk}}_{\textrm{Ca}}\right)+\Gamma^i_{\textrm{Mn,O}}\left(\Delta\mu_{\textrm{O}}+\frac{1}{2}\left(E^{\textrm{gas}}_{\textrm{O$_{2}$}}+F^{\textrm{vib,gas}}_{\textrm{O$_{2}$}}\right)\right)\right].
\end{eqnarray}
The same approach including the vibrational free energy is applied 
to Eq.\ (4), whose expression is rewritten introducing the formation energy, 
$\Delta$$E$$^{\textrm{bulk}}_{\textrm{f}}$, as follows,
\begin{eqnarray}
\Delta\mu_{\textrm{Ca}}+\Delta\mu_{\textrm{Mn}}+3\Delta\mu_{\textrm{O}}&=&\Delta{E}^{\textrm{bulk}}_{\textrm{f,CaMnO$_{3}$}}+F^{\textrm{vib,bulk}}_{\textrm{CaMnO$_{3}$}}.
\end{eqnarray}
This equation is then used to rearrange all the bulk stability 
boundary inequalities, Eq.\ (6-17), so that only $\mu_{\textrm{Ca}}$ and 
$\mu_{\textrm{O}}$ are kept as variables.
For instance, Eq.\ (15) becomes
\begin{eqnarray}
\Delta\mu_{\textrm{Ca}}+\Delta\mu_{\textrm{O}}&\le&\Delta{E}^{\textrm{bulk}}_{\textrm{f,Ca$_{2}$MnO$_{4}$}}
-\Delta{E}^{\textrm{bulk}}_{\textrm{f,CaMnO$_{3}$}} \nonumber \\
                                              &&   +F^{\textrm{vib,bulk}}_{\textrm{Ca$_{2}$MnO$_{4}$}}-
F^{\textrm{vib,bulk}}_{\textrm{CaMnO$_{3}$}}.
\end{eqnarray}
The results from the DFT formation energies, 
$\Delta$$E$$^{\textrm{bulk}}_{\textrm{f}}$, for all the 
binary and ternary sub-phases are compared to experimental quantities 
in Table 2.
The formation energies are slightly overestimated, which is a
well-known shortcoming of DFT~\cite{Kotomin08p4644}.
The surface phase diagram dependence on \emph{T} and \emph{p}(O$_{2}$) 
provides insightful physical interpretations of the experimental range of 
conditions for the surface stability.
The $\Delta\mu_{\textrm{O}}$ can be directly related to \emph{T} and 
\emph{p}(O$_{2}$) by the ideal gas approximation, since at equilibrium 
the $\mu_{\textrm{O}}$ in CaMnO$_{3}$ bulk is equal to the chemical potential of oxygen 
gas in the environment, 
$\mu_{\textrm{O}}$=$\frac{1}{2}$$\mu^{\textrm{gas}}_{\textrm{O$_{2}$}}$(\emph{T},\emph{p}). 
This leads to 
\begin{eqnarray}
\Delta\mu_{\textrm{O}}(T,p) = \Delta\mu_{\textrm{O}}\left(T,p^{\textrm{0}}\right)+\frac{1}{2}kT\ln\left(\frac{p}{p^{\textrm{0}}}\right)
\end{eqnarray}
where $\Delta\mu$$_{\textrm{O}}$(\emph{T},\emph{p}$^0$)=$\frac{1}{2}$$\Delta$\emph{G}$^{\textrm{gas}}_{\textrm{O$_{2}$}}$(\emph{T},\emph{p}$^0$).
Thermodynamic data from the NIST-JANAF thermochemical tables~\cite{Chase98p1}
are used to determine the values of $\Delta\mu$$_{\textrm{O}}$(\emph{T},\emph{p}$^0$) 
by selecting the reference state for 
$\frac{1}{2}$$\Delta$\emph{G}$^{\textrm{gas}}_{\textrm{O$_{2}$}}$(\emph{T},\emph{p}$^0$)
extrapolated to \emph{T}=0 K and \emph{p}(O$_{2}$)=1 atm. 
Then, the values obtained are plotted for a temperature range of 
0\textrm{--}3000 K at different \emph{p}(O$_{2}$) to correlate the stable surface 
phases with controlled environmental conditions.

\section*{\label{sec:level1} III. Results and Discussion}

\subsection*{\label{sec:level2} A. Magnetic Surface Reconstruction }

\begin{figure}
\centering
\includegraphics[width=0.7\textwidth, angle=90]{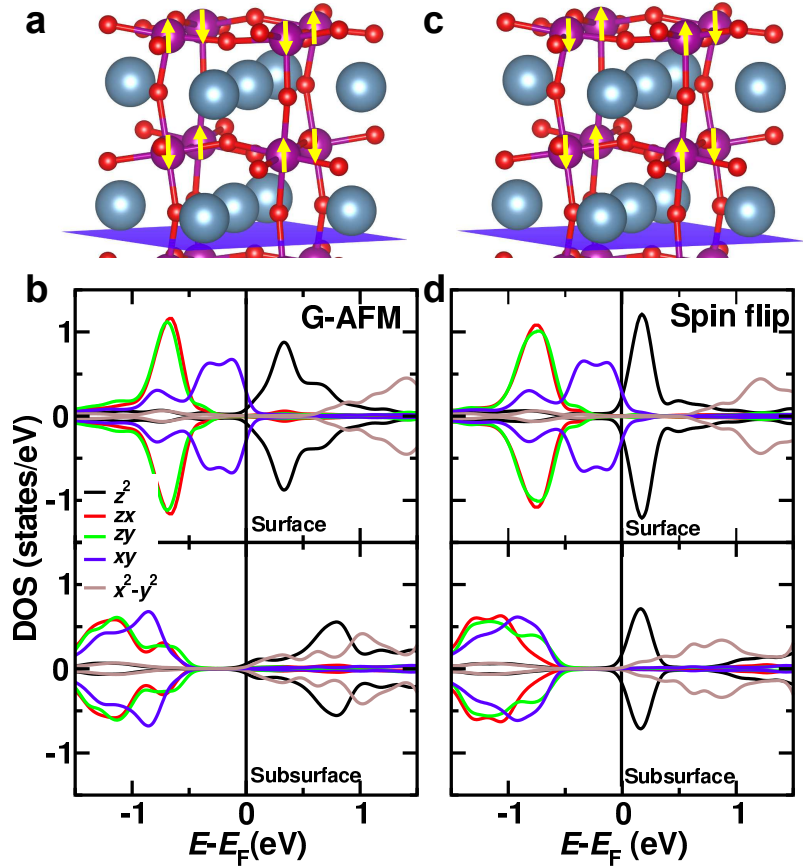}
\caption{Spin-flip magnetic reconstruction at the stoichiometric MnO$_{2}$-terminated 
surface. The atomic, electronic and magnetic structure of the stoichiometric 
MnO$_{2}$-terminated surface adopting (a,b) the bulk magnetic configuration, G-AFM, 
and (c,d) an energetically more favorable magnetic reconstruction where the spins of 
the surface Mn atoms are flipped. The magnetic spins are represented by the yellow arrows. 
The orbital-projected densities of states (DOS) for surface and subsurface 
Mn \emph{d}-orbitals of the (b) G-AFM, and (d) spin-flip magnetic structures show 
different energetic alignment of the 3\emph{d$_{z^{2}}$} orbital between surface 
and subsurface states.}
\label{Fig.2}
\end{figure}

Depending on their surface composition, some of the MnO$_{2}$-based 
slabs undergo a magnetic spin flip reconstruction at the surface~\cite{Filippetti99p4184} 
that is energetically favorable, leading to lower total energy and greater surface 
stability.
This magnetic spin flip reconstruction refers to a surface with G-AFM bulk order, but 
where the Mn spins in the surface layer align with the spins of the 
Mn in the second layer, generating a bilayer with C-type AFM alignment, as shown in Fig.\ \ref{Fig.2}(a,c).
This occurs when the surface breaks the degeneracy of \emph{e$_{g}$}  
and \emph{t$_{2g}$} states, as the Mn cage is changed from octahedral 
to square pyramidal.
This leads to the partial occupation of the \emph{e$_{g}$} 
\emph{d$_{z^{2}}$} bands at the surface, which favors the 
FM alignment of spins between the surface and subsurface 
MnO$_{2}$ layers.
In bulk CaMnO$_{3}$, the superexchange magnetic interactions between the 
filled \emph{t$_{2g}$} shells and the formally empty \emph{e$_{g}$} shells 
determine the G-AFM ordering.
However, this G-AFM ordering does not extend to the (001) MnO$_{2}$-terminated 
surfaces, since double-exchange governs the magnetic interaction between Mn subsurface and surface.
The Mn 3\emph{d} orbital-projected density of states (DOS) of both
G-AFM and spin flip surface and subsurface layers of the MnO$_{2}$-terminated 
stoichiometric surface are shown in Fig.\ \ref{Fig.2}(b,d).
The surface states of both G-AFM and spin-flip show the broken 
degeneracy of \emph{e$_{g}$} $\rightarrow$ \emph{d$_{x^{2}-y^{2}}$}, 
\emph{d$_{z^{2}}$} and \emph{t$_{2g}$} $\rightarrow$ \emph{d$_{xy}$}, 
(\emph{d$_{xz}$},\emph{d$_{yz}$}), as expected for square pyramidal.
The G-AFM subsurface states have degenerate \emph{e$_{g}$} and \emph{t$_{2g}$} 
shells; however, the spin-flip subsurface \emph{e$_{g}$} degeneracy is broken, 
as the \emph{d$_{z^{2}}$} is favorably aligned with the \emph{d$_{z^{2}}$} 
from its neighboring surface state.
The alignment of the Mn spins on surface and subsurface layers is 
induced by a double-exchange process, which is enabled by the occupation of 
the \emph{d$_{z^{2}}$} orbital, lowering the energy of MnO$_{2}$-terminated surfaces 
with the spin flip arrangement. 
Our DFT computed stoichiometric MnO$_{2}$-surface yields a spin flip 
structure as its magnetic ground state, with an energy lowering relative to G-AFM of 
15 meV per primitive surface cell.  

Based on the prevalence of the energetically favorable magnetic spin flip reconstruction on 
the stoichiometric MnO$_{2}$-terminated surface, it is important to study the effect of magnetic 
reconstruction on surface stability. 
All the MnO$_{2}$-based surface reconstructions were relaxed with both G-AFM and spin flip magnetic 
ordering.
Different magnetic configurations were explored for CaO-terminated surfaces with Mn adatoms as well as 
MnO$_{2}$ terminated surfaces with Mn$_{x}$O$_{y}$ adatom reconstructions (See Supplementary Material). 
Half of the thermodynamically stable MnO$_{2}$-based surfaces are magnetic spin flip configurations, as 
shown in Fig.\ \ref{Fig.3} (surface phases with red grid lines).   
For surface reconstructions with only one extra Mn adatom, the spin arrangement that makes a local 
FM alignment between the surface Mn adatom and the nearest Mn atom is more favorable.
For example, the local FM alignment between the Mn surface adatom and the Mn on the 
subsurface layer in CaO-0.5Ca+0.5Mn, which is within the bulk stability region, is 12 meV per 
primitive surface cell more stable than a local AFM alignment (See Supplementary Material).

\subsection*{\label{sec:level2} B. Surface Phase Diagram}

\begin{figure*}[h]
\centering
\includegraphics[width=0.35\textwidth, angle=90]{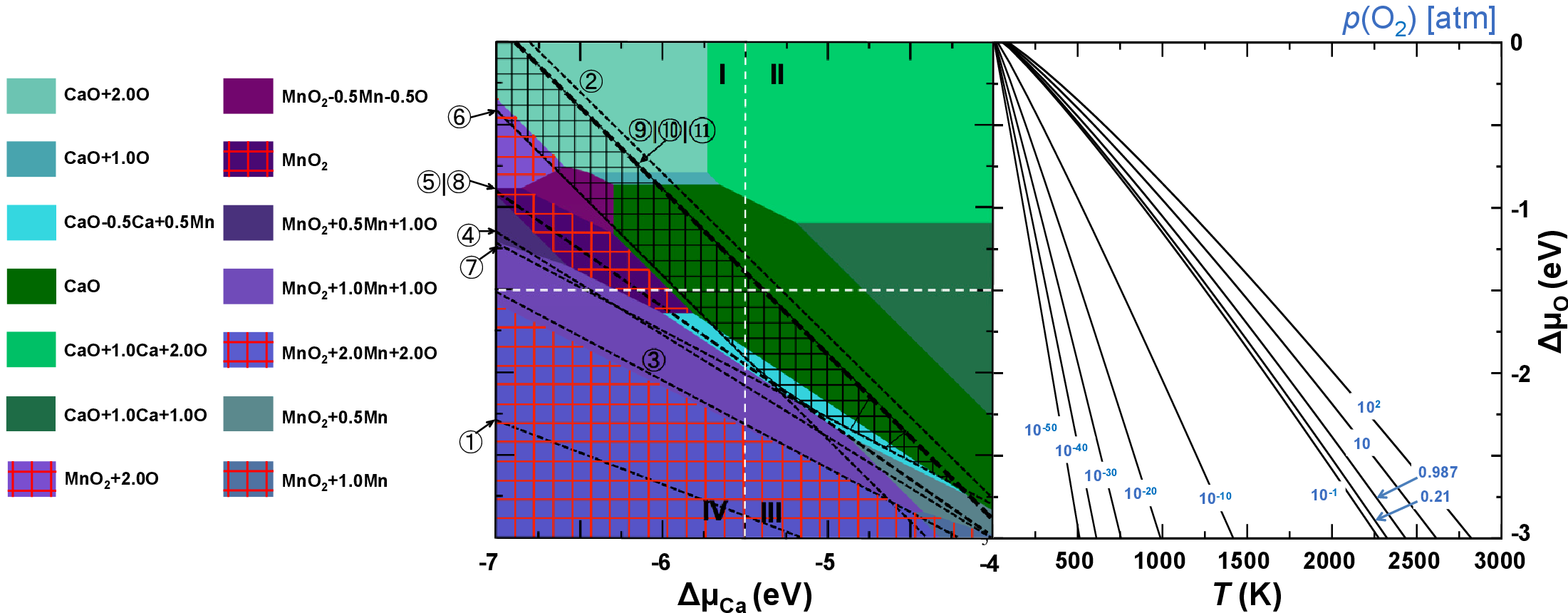}
\caption{Surface phase diagram of CaMnO$_{3}$ with Mn metallic, binary, and 
ternary sub-phase bulk stability boundaries at \emph{T}=0 K denoted as 
follows (1)Mn, (2)CaO, (3)MnO, (4)Mn$_{3}$O$_{4}$, 
(5)Mn$_{2}$O$_{3}$, (6)MnO$_{2}$, (7)CaMn$_{2}$O$_{4}$, 
(8)CaMn$_{7}$O$_{12}$, (9)Ca$_{2}$MnO$_{4}$, 
(10)Ca$_{3}$Mn$_{2}$O$_{7}$, and (11)Ca$_{4}$Mn$_{3}$O$_{10}$. The 
enclosed area shaded by black grid lines represents the bulk stability region.  
The phase diagram is divided into four regions, with dashed white lines to group 
phases with thermodynamic stability at similar condition ranges of 
$\Delta\mu_{\textrm{O}}$, $\Delta\mu_{\textrm{Ca}}$, and $\Delta\mu_{\textrm{Mn}}$.  
The color-coded legend for each of the surface phases is on the left panel. 
The phases with red grid lines represent surfaces with spin-flip magnetic 
reconstructions. The right panel shows the \emph{T} and \emph{p}(O$_{2}$) 
dependence of $\Delta\mu_{\textrm{O}}$. }
\label{Fig.3}
\end{figure*}

The computed surface phase diagram of CaMnO$_{3}$, generated after calculating the surface free energy of the 
atomic and magnetic lowest energy configuration of each surface phase, is shown in Fig.\ \ref{Fig.3}.
This diagram predicts a total of fourteen thermodynamically stable 
surface phases in the given range of $\Delta\mu$$_{\textrm{Ca}}$ and 
$\Delta\mu$$_{\textrm{O}}$.
The phase diagram can be divided into four different quadrants based 
on conditions of $\Delta\mu_{\textrm{O}}$, $\Delta\mu_{\textrm{Ca}}$ 
and $\Delta\mu_{\textrm{Mn}}$.
Even though $\Delta\mu_{\textrm{Mn}}$ is not directly plotted, its 
value can always be deduced from Eq.\ (25).
The upper left quadrant (I) shows oxygen rich and cation poor 
conditions; therefore, the favored phases are terminations 
with excess oxygen adatoms, denoted as MnO$_{2}$+2.0O, CaO+2.0O and 
CaO+1.0O.
This quadrant also shows stability of the stoichiometric CaO- and MnO$_{2}$-terminated 
surfaces as well as MnO$_{2}$-based reconstructions, MnO$_{2}$-0.5Mn-0.5O and MnO$_{2}$+0.5Mn+1.0O.
In quadrant (II), both O and Ca are plentiful, leading to the absence of 
MnO$_{2}$-terminated stable surfaces and the presence of Ca adatoms in stable 
CaO-based surfaces, CaO+1.0Ca+2.0O and CaO+1.0Ca+1.0O.
The bottom right quadrant (III) involves oxygen-poor but relatively 
Ca-rich conditions.
This region is characterized by stable surfaces with MnO$_{2}$ 
termination, denoting that even though Ca is at high concentrations, 
MnO$_{2}$-terminated surfaces are more favorable at oxygen poor 
conditions, clearly indicated by a surface phase where a Ca vacancy is 
filled by a Mn adatom, CaO-0.5Ca+0.5Mn.
In the left bottom quadrant (IV), Mn is at its richest 
condition since both O and Ca are at poor levels.
This promotes the large stable region for both MnO$_{2}$+1.0Mn+1.0O and MnO$_{2}$+2.0Mn+2.0O.
A boundary line from the top left to mid-bottom right corners 
of the diagram delimits the regions of CaO- (upper) 
and MnO$_{2}$- (lower) terminated surfaces.
Six different CaO-terminated surfaces are found, many of them covering 
a significant area of chemical potential space.
On the other hand, MnO$_{2}$-type phases generally show smaller stable 
regions with a higher variety of phases, which can be explained by the 
multivalency of Mn. 
For instance, the predicted path from MnO$_{2}$+2.0Mn+2.0O to MnO$_{2}$+2.0O 
consists of a complex series of transitions.
These transitions occur by increasing the $\Delta\mu_{\textrm{O}}$ 
from -3.0 eV to -0.5 eV while keeping constant the 
$\Delta\mu_{\textrm{Ca}}$ at $\approx$-7.0 eV.
The changes in $\Delta\mu_{\textrm{O}}$ can be controlled 
experimentally by varying \emph{T} or \emph{p}(O$_{2}$).
For example, these surface phase transitions are predicted to occur at a 
constant \emph{p}(O$_{2}$) of 10$^{-10}$ atm, as the temperature is decreased 
from $\approx$1400 K to $\approx$400 K.
Starting with the surface initially at $\approx$1400 K in 
the MnO$_{2}$+2.0Mn+2.0O phase, as the temperature is reduced to $\approx$750 K, the 
desorption of oxygen atoms is thermodynamically favorable, leading to 
the surface reconstruction MnO$_{2}$+1.0Mn+1.0O. 
Reducing the temperature further to $\approx$600 K should cause the surface to 
reconstruct into MnO$_{2}$+0.5Mn+1.0O and even co-exist with stoichiometric MnO$_{2}$-terminated 
surface. 
Once the temperature reaches $\approx$400 K, the surface phase 
transition to MnO$_{2}$+2.0O is thermodynamically favored.
The slopes of the boundaries between phases directly relate to the changes 
in numbers of surface atoms of each species.
The CaO+2.0O surface phase changes to CaO+1.0Ca+2.0O at a
vertical boundary of $\Delta\mu_{\textrm{Ca}}$ $\approx$-5.75 eV 
via adsorption of Ca cations.
Similarly, horizontal lines at $\Delta\mu_{\textrm{O}}$ $\approx$-0.8 
eV and $\approx$-0.9 eV denotes its transition to CaO+1.0O and stoichiometric CaO, 
respectively, by the removal of oxygen adatoms.
Between these extremes, the slope of the CaO to CaO+1.0Ca+2.0O 
transition, as indicated in Fig.\ \ref{Fig.3}, is -$\frac{1}{2}$, 
suggesting the addition of 2 Ca cations and 4 oxygen atoms per cell 
since the transition line will be crossed by moving toward Ca- and O-rich conditions.
Taking into account the relationship shown in Eq.\ (25), the slope of the 
boundaries between surface phases can then be defined as 
$\frac{(\emph{a}-\emph{b})}{(3\emph{b}-\emph{c})}$, where \emph{a}, \emph{b}, 
and \emph{c} represent the net change in number of Ca, Mn, and O atoms, respectively, 
between phases.
For example, the removal of 2 Mn atoms and the adsorption of 2 oxygen atoms 
per cell results in the transition of surface phase MnO$_{2}$+2.0Mn+2.0O to 
MnO$_{2}$+1.0Mn+1.0O. 
This is in agreement with the boundary for this phase transition having slope of -$\frac{1}{4}$.

\subsubsection*{\label{sec:level3}  Stability Region}

The surface phase diagram shows a series of stable surfaces, but 
only a few of them are within the bulk stability region.
The ternary sub-phases illustrated in Eqs.\ 13-17 are included 
in the definition of the stability region since an extensive
experimental examination of the phase relations of Ca-Mn-O bulk 
systems~\cite{Horowitz78p1359, Balakirev06pS49} found the 
co-existence lines for the CaMnO$_{3}$ phase with these ternary sub-phases, 
and correspondingly narrower stability boundaries for CaMnO$_{3}$.  
The boundary of bulk phase stability is defined by Ca$_{2}$MnO$_{4}$, 
Ca$_{3}$Mn$_{2}$O$_{7}$, and Ca$_{4}$Mn$_{3}$O$_{10}$ on the top and 
Mn$_{2}$O$_{3}$, MnO$_{2}$, CaMn$_{2}$O$_{4}$, and CaMn$_{7}$O$_{12}$ 
on the bottom (lines 9-11 and 5-8, respectively, shown in Fig.\ \ref{Fig.3}).
A confirmation of the importance of including the ternary sub-phases 
is the ability of this model to describe the decomposition of bulk 
CaMnO$_{3}$.  
Experimentally, decomposition of CaMnO$_{3}$ occurs at $\approx$1700 K and 
0.21 atm O$_{2}$~\cite{Horowitz78p1359}. 
Using the DFT-generated stability region computed at 0 K, this 
decomposition temperature is predicted to be $\approx$2100 K at 
0.21 atm by considering the decomposition conditions at the low 
$\Delta\mu_{\textrm{O}}$ end of the bulk stability region.
However, if boundaries 7-11 are neglected, then the stability 
region is limited by MnO$_{2}$, Mn$_{2}$O$_{3}$ and Mn$_{3}$O$_{4}$ on the 
bottom and bordered by CaO on the top (lines 6, 5, 4, and 2, respectively, in 
Fig.\ \ref{Fig.2}), artificially increasing the computed decomposition 
temperature to over 2500 K at 0.21 atm.
These ternary sub-phases include the Ruddlesden-Popper phases 
CaO(CaMnO$_{3}$)$_{n}$, which are perovskite structures with 
double CaO layer (001) planes with \emph{n} layers of CaMnO$_{3}$ 
in between~\cite{Fawcett98p3643}.
Naturally, the surface reconstruction with CaO double layer,
CaO+1.0Ca+1.0O, is found on the side where the Ruddlesden-Popper phases 
CaO(CaMnO$_{3}$)$_{n}$ limit the stability of the CaMnO$_{3}$ bulk.
The CaMn$_{2}$O$_{4}$ phase is a marokite structure~\cite{Ling01p167}, 
where the MnO$_{6}$ octahedra share edges and corners.
This network of octahedra leads to channels along the \emph{a}-axis which 
provide Ca atoms with 8-coordinated sites. 
It has been found that this marokite-type phase shows potential photocatalytic 
activity~\cite{Wu11p107101}. 
This bulk phase borders the surface phase with a MnO$_{2}$-terminated 
layer that has an additional MnO layer, MnO$_{2}$+1.0Mn+1.0O.
CaMn$_{7}$O$_{12}$ is a complex distorted double perovskite, whose 
chemical formula is formally written as 
(CaMn$^{+3}_{3}$)(Mn$^{+3}_{3}$Mn$^{+4}$)O$_{12}$.
This perovskite-derived phase consists of $\frac{1}{4}$ of the \emph{A}-sites 
occupied by Ca and the rest by Mn$^{+3}$, while the \emph{B}-site charges are 
arranged via charge ordering below 440 K~\cite{Bochu80p133}.
All these ternary sub-phases are stable for some range of chemical 
potentials~\cite{Horowitz78p1359, Balakirev06pS49}, defining 
realistic stability boundaries for the surface of CaMnO$_{3}$.  

\subsubsection*{\label{sec:level3} Temperature Dependence}

\begin{figure*}
\centering
\includegraphics[width=0.55\textwidth, angle = 90]{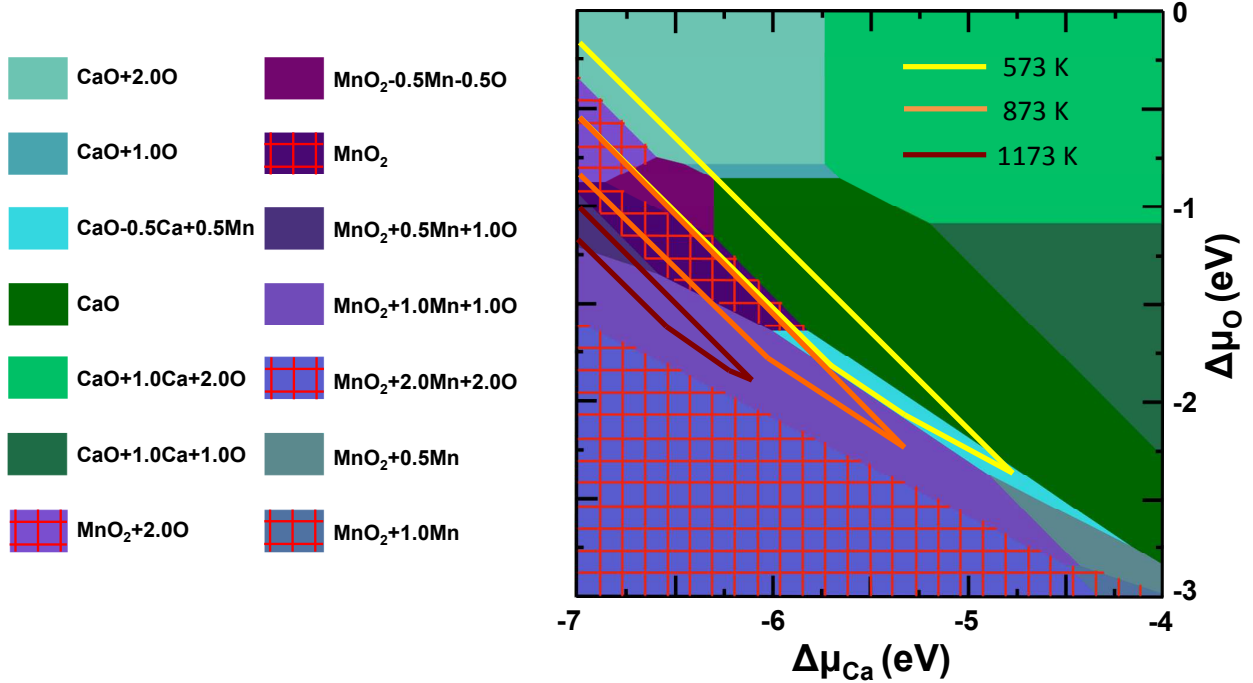}
\caption{Temperature dependence of the bulk stability region within CaMnO$_{3}$ surface phase diagram. Stability limits at different temperatures 573 K (yellow), 873 K (orange) and 1173 K (dark red) are plotted, indicating the thermodynamic stability of MnO$_{2}$-terminated surface reconstructions.}
\label{Fig.4}
\end{figure*}

The bulk stability region described above is based on DFT 
calculations at 0 K enclosing the CaO+2.0O, CaO+1.0O, CaO, CaO-0.5Ca+0.5Mn, 
MnO$_{2}$+2.0O, and MnO$_{2}$-0.5Mn-0.5O phases. 
A more accurate description of the stability region requires
the inclusion of the vibrational free energy.
This is significant, since it determines how the stability region 
changes as a function of temperature.
The bulk stability region is recalculated incorporating the bulk vibrational 
free energy at 573 K, 873 K and 1173 K, shown in Fig.\ \ref{Fig.4}.
These temperatures are selected since they are within the range of 
experimental temperature conditions used in the synthesis and pulsed 
laser deposition of the CaMnO$_{3}$ crystal~\cite{Melo01p915} 
and film, as well as in the annealing process for surface characterization 
techniques.
These temperatures shift the stability region such that only MnO$_{2}$-based 
surfaces are favored.
As temperature increases, the stability region shifts towards Ca-poor and O-poor 
conditions, in agreement with previous studies on other surfaces~\cite{Lee11p115418, Kolpak08p036102}. 
As the stability region moves, it encloses other phases that are 
predicted to be stable at those temperatures.
At 573 K, the surface phases within the bulk stability region are the same as in Fig.\ \ref{Fig.3}, with the addition of the stoichiometric MnO$_{2}$-terminated surface. 
Increasing the temperature by 300 K leads to the stability of MnO$_{2}$+1.0Mn+1.0O surface phase. 
At 1173 K, the phases within the bulk stability region include the MnO$_{2}$+0.5Mn+1.0O phase.
The clear trend is that the surface structures with MnO$_{2}$ 
terminations are more favorable at higher temperatures.
Changes in the shape of the stability region reflect 
how binary and ternary sub-phase boundaries shift as the vibrational 
free energy at various temperatures is included.
Overall, the bulk stability region tends to shrink as the higher temperature 
contributions are included.
This is in good agreement with previous reports, since it is experimentally observed 
that CaMnO$_{3}$ crystal decomposes at temperatures higher than $\approx$1700 K~\cite{Horowitz78p1359} 
and our theoretical bulk stability region vanishes around the same temperature. 

\section*{\label{sec:level1} IV. Summary and Conclusions}
The energies of series of CaO- and MnO$_{2}$-terminated CaMnO$_{3}$ (001) 
surfaces with different combinations of vacancies, adatoms and 
additional layers are reported using \emph{ab initio} thermodynamics 
to theoretically predict the surface phase diagram.
The predicted surface phase diagram has surface structures 
for both terminations, with stability ranges specified by 
$\Delta\mu_{\textrm{O}}$ and $\Delta\mu_{\textrm{Ca}}$.
The $\Delta\mu_{\textrm{O}}$ is related to \emph{T} and \emph{p}(O$_{2}$), 
linking surface stability with experimental conditions.
The chemical potential region for which CaMnO$_{3}$ is the only bulk 
phase is bounded by coexistence lines with the Ruddlesden-Popper phases
CaO(CaMnO$_{3}$)$_{\emph{n}}$ (where \emph{n}=1, 2, and 3), as well as 
Mn$_{2}$O$_{3}$, MnO$_{2}$, CaMn$_{2}$O$_{4}$, and CaMn$_{7}$O$_{12}$.
This shows the relevance of incorporating the ternary subphases from 
the experimental Ca-Mn-O compositional phase 
diagram~\cite{Horowitz78p1359, Balakirev06pS49}.
The surface phases that are favored within this bulk stability region 
change once the vibrational free energy at different temperatures is included.
The stability region shifts to lower $\Delta\mu_{\textrm{Ca}}$ for higher 
\emph{T}, leading to the predominance of complex MnO$_{2}$-based 
surfaces at \emph{T} above 800 K.
Thermodynamically stable phases include potentially reactive surfaces, 
MnO$_{2}$+1.0Mn+1.0O, since it shows two unique Mn sites based on coordination.   
The MnO$_{2}$-based surfaces show intriguing atomic and magnetic structures which are comprehensively 
explored by computing the magnetic ground state and analyzing its thermodynamic stability.
Some of the MnO$_{2}$-terminated surfaces are energetically more 
favorable when their surface and subsurface spins are aligned.
This occurs when the filling of the \emph{d$_{z^{2}}$} orbital causes a more favorable double-exchange 
interaction between the subsurface and surface Mn atoms.
The atomic composition of the surface affects this magnetic reconstruction; for some Mn$_{x}$O$_{y}$ 
reconstructions, the surface reconstructs, rearranging the magnetic interactions in the system.
However, this magnetic spin flip reconstruction prevails for key stable surfaces such as 
stoichiometric MnO$_{2}$-terminated and MnO$_{2}$+2.0O.     
This study addresses structures with ($\sqrt{2}\times\sqrt{2}$)\emph{R}45$^{\circ}$ 
surface symmetry, leading to ordered surfaces due to 
periodic boundary conditions.
Admittedly, real surfaces are more complex and disordered; however, 
by taking into account the chemical and thermal contributions to the 
surface stability and exploring a wide range of possible surface 
structures, the predicted phase diagram provides an acceptable foundation 
for the analysis of experimental data on this surface.
Consequently, this predicted surface phase diagram could lead to 
experimental studies performing low-energy electron diffraction 
and scanning tunneling microscopy to validate the stable surface 
reconstructions and further explore the surface evolution.

\section*{\label{sec:level1} Acknowledgments}

D.S.-G. was supported by the Office of Naval Research under Grant No.\ N00014-11-1-0664.
C.-W.L. was supported by the US Department of Energy Office of Basic Energy Sciences 
under Grant No.\ DE-FG02-07ER15920.
D.Y. was supported by the National Science Foundation under Grant No.\ DMR-1124696. 
A.M.R. was supported by the National Science Foundation under Grant No.\ CMMI-1334241.
Computational support was provided by the High-Performance Computing Modernization 
Office of the US Department of Defense, and the National Energy Research Scientific 
Computing Center of the US Department of Energy.

\bibliography{rappecites} 

%merlin.mbs apsrev4-1.bst 2010-07-25 4.21a (PWD, AO, DPC) hacked
%Control: key (0)
%Control: author (8) initials jnrlst
%Control: editor formatted (1) identically to author
%Control: production of article title (-1) disabled
%Control: page (0) single
%Control: year (1) truncated
%Control: production of eprint (0) enabled
\begin{thebibliography}{47}%
\makeatletter
\providecommand \@ifxundefined [1]{%
 \@ifx{#1\undefined}
}%
\providecommand \@ifnum [1]{%
 \ifnum #1\expandafter \@firstoftwo
 \else \expandafter \@secondoftwo
 \fi
}%
\providecommand \@ifx [1]{%
 \ifx #1\expandafter \@firstoftwo
 \else \expandafter \@secondoftwo
 \fi
}%
\providecommand \natexlab [1]{#1}%
\providecommand \enquote  [1]{``#1''}%
\providecommand \bibnamefont  [1]{#1}%
\providecommand \bibfnamefont [1]{#1}%
\providecommand \citenamefont [1]{#1}%
\providecommand \href@noop [0]{\@secondoftwo}%
\providecommand \href [0]{\begingroup \@sanitize@url \@href}%
\providecommand \@href[1]{\@@startlink{#1}\@@href}%
\providecommand \@@href[1]{\endgroup#1\@@endlink}%
\providecommand \@sanitize@url [0]{\catcode `\\12\catcode `\$12\catcode
  `\&12\catcode `\#12\catcode `\^12\catcode `\_12\catcode `\%12\relax}%
\providecommand \@@startlink[1]{}%
\providecommand \@@endlink[0]{}%
\providecommand \url  [0]{\begingroup\@sanitize@url \@url }%
\providecommand \@url [1]{\endgroup\@href {#1}{\urlprefix }}%
\providecommand \urlprefix  [0]{URL }%
\providecommand \Eprint [0]{\href }%
\providecommand \doibase [0]{http://dx.doi.org/}%
\providecommand \selectlanguage [0]{\@gobble}%
\providecommand \bibinfo  [0]{\@secondoftwo}%
\providecommand \bibfield  [0]{\@secondoftwo}%
\providecommand \translation [1]{[#1]}%
\providecommand \BibitemOpen [0]{}%
\providecommand \bibitemStop [0]{}%
\providecommand \bibitemNoStop [0]{.\EOS\space}%
\providecommand \EOS [0]{\spacefactor3000\relax}%
\providecommand \BibitemShut  [1]{\csname bibitem#1\endcsname}%
\let\auto@bib@innerbib\@empty
%</preamble>
\bibitem [{\citenamefont {Bhattacharjee}\ \emph {et~al.}(2009)\citenamefont
  {Bhattacharjee}, \citenamefont {Bousquet},\ and\ \citenamefont
  {Ghosez}}]{Bhattacharjee09p117602}%
  \BibitemOpen
  \bibfield  {author} {\bibinfo {author} {\bibfnamefont {S.}~\bibnamefont
  {Bhattacharjee}}, \bibinfo {author} {\bibfnamefont {E.}~\bibnamefont
  {Bousquet}}, \ and\ \bibinfo {author} {\bibfnamefont {P.}~\bibnamefont
  {Ghosez}},\ }\href {\doibase 10.1103/PhysRevLett.102.117602} {\bibfield
  {journal} {\bibinfo  {journal} {Phys. Rev. Lett.}\ }\textbf {\bibinfo
  {volume} {102}},\ \bibinfo {pages} {117602} (\bibinfo {year}
  {2009})}\BibitemShut {NoStop}%
\bibitem [{\citenamefont {Wang}\ \emph {et~al.}(2012)\citenamefont {Wang},
  \citenamefont {He},\ and\ \citenamefont {Wu}}]{Wang12p17005}%
  \BibitemOpen
  \bibfield  {author} {\bibinfo {author} {\bibfnamefont {H.}~\bibnamefont
  {Wang}}, \bibinfo {author} {\bibfnamefont {L.}~\bibnamefont {He}}, \ and\
  \bibinfo {author} {\bibfnamefont {X.}~\bibnamefont {Wu}},\ }\href
  {http://stacks.iop.org/0295-5075/100/i=1/a=17005} {\bibfield  {journal}
  {\bibinfo  {journal} {EPL (Europhysics Letters)}\ }\textbf {\bibinfo {volume}
  {100}},\ \bibinfo {pages} {17005} (\bibinfo {year} {2012})}\BibitemShut
  {NoStop}%
\bibitem [{\citenamefont {Urata}\ \emph {et~al.}(2007)\citenamefont {Urata},
  \citenamefont {Funahashi1}, \citenamefont {Mihara}, \citenamefont {Kosuga},
  \citenamefont {Sodeoka},\ and\ \citenamefont {Tanaka}}]{Urata07p535}%
  \BibitemOpen
  \bibfield  {author} {\bibinfo {author} {\bibfnamefont {S.}~\bibnamefont
  {Urata}}, \bibinfo {author} {\bibfnamefont {R.}~\bibnamefont {Funahashi1}},
  \bibinfo {author} {\bibfnamefont {T.}~\bibnamefont {Mihara}}, \bibinfo
  {author} {\bibfnamefont {A.}~\bibnamefont {Kosuga}}, \bibinfo {author}
  {\bibfnamefont {S.}~\bibnamefont {Sodeoka}}, \ and\ \bibinfo {author}
  {\bibfnamefont {T.}~\bibnamefont {Tanaka}},\ }\href@noop {} {\bibfield
  {journal} {\bibinfo  {journal} {International Journal of Applied Ceramic
  Technology}\ }\textbf {\bibinfo {volume} {4}},\ \bibinfo {pages} {535}
  (\bibinfo {year} {2007})}\BibitemShut {NoStop}%
\bibitem [{\citenamefont {Koumoto}\ \emph {et~al.}(2010)\citenamefont
  {Koumoto}, \citenamefont {Wang}, \citenamefont {Zhang}, \citenamefont
  {Kosuga},\ and\ \citenamefont {Funahashi}}]{Koumoto10p363}%
  \BibitemOpen
  \bibfield  {author} {\bibinfo {author} {\bibfnamefont {K.}~\bibnamefont
  {Koumoto}}, \bibinfo {author} {\bibfnamefont {Y.}~\bibnamefont {Wang}},
  \bibinfo {author} {\bibfnamefont {R.}~\bibnamefont {Zhang}}, \bibinfo
  {author} {\bibfnamefont {A.}~\bibnamefont {Kosuga}}, \ and\ \bibinfo {author}
  {\bibfnamefont {R.}~\bibnamefont {Funahashi}},\ }\href@noop {} {\bibfield
  {journal} {\bibinfo  {journal} {Annual Review of Materials Research}\
  }\textbf {\bibinfo {volume} {40}},\ \bibinfo {pages} {363} (\bibinfo {year}
  {2010})}\BibitemShut {NoStop}%
\bibitem [{\citenamefont {Zeng}\ \emph {et~al.}(1999)\citenamefont {Zeng},
  \citenamefont {Greenblatt},\ and\ \citenamefont {Croft}}]{Zeng99p8784}%
  \BibitemOpen
  \bibfield  {author} {\bibinfo {author} {\bibfnamefont {Z.}~\bibnamefont
  {Zeng}}, \bibinfo {author} {\bibfnamefont {M.}~\bibnamefont {Greenblatt}}, \
  and\ \bibinfo {author} {\bibfnamefont {M.}~\bibnamefont {Croft}},\ }\href
  {\doibase 10.1103/PhysRevB.59.8784} {\bibfield  {journal} {\bibinfo
  {journal} {Phys. Rev. B}\ }\textbf {\bibinfo {volume} {59}},\ \bibinfo
  {pages} {8784} (\bibinfo {year} {1999})}\BibitemShut {NoStop}%
\bibitem [{\citenamefont {Kawashima}\ \emph {et~al.}(2008)\citenamefont
  {Kawashima}, \citenamefont {Matsubara},\ and\ \citenamefont
  {Honda}}]{Kawashima08p3439}%
  \BibitemOpen
  \bibfield  {author} {\bibinfo {author} {\bibfnamefont {A.}~\bibnamefont
  {Kawashima}}, \bibinfo {author} {\bibfnamefont {K.}~\bibnamefont
  {Matsubara}}, \ and\ \bibinfo {author} {\bibfnamefont {K.}~\bibnamefont
  {Honda}},\ }\href@noop {} {\bibfield  {journal} {\bibinfo  {journal}
  {Bioresource Technology}\ }\textbf {\bibinfo {volume} {99}},\ \bibinfo
  {pages} {3439} (\bibinfo {year} {2008})}\BibitemShut {NoStop}%
\bibitem [{\citenamefont {Najafpour}\ \emph {et~al.}(2012)\citenamefont
  {Najafpour}, \citenamefont {Pashaei},\ and\ \citenamefont
  {Nayeri}}]{Najafpour12p1}%
  \BibitemOpen
  \bibfield  {author} {\bibinfo {author} {\bibfnamefont {M.~M.}\ \bibnamefont
  {Najafpour}}, \bibinfo {author} {\bibfnamefont {B.}~\bibnamefont {Pashaei}},
  \ and\ \bibinfo {author} {\bibfnamefont {S.}~\bibnamefont {Nayeri}},\ }\href
  {\doibase 10.1039/C2DT12189A} {\bibfield  {journal} {\bibinfo  {journal}
  {Dalton Trans.}\ }\textbf {\bibinfo {volume} {41}},\ \bibinfo {pages} {4799}
  (\bibinfo {year} {2012})}\BibitemShut {NoStop}%
\bibitem [{\citenamefont {Filippetti}\ and\ \citenamefont
  {Pickett}(1999)}]{Filippetti99p4184}%
  \BibitemOpen
  \bibfield  {author} {\bibinfo {author} {\bibfnamefont {A.}~\bibnamefont
  {Filippetti}}\ and\ \bibinfo {author} {\bibfnamefont {W.~E.}\ \bibnamefont
  {Pickett}},\ }\href@noop {} {\bibfield  {journal} {\bibinfo  {journal} {Phys.
  Rev. Lett.}\ }\textbf {\bibinfo {volume} {83}},\ \bibinfo {pages} {4184}
  (\bibinfo {year} {1999})}\BibitemShut {NoStop}%
\bibitem [{\citenamefont {Filippetti}\ and\ \citenamefont
  {Pickett}(2000)}]{Filippetti00p11571}%
  \BibitemOpen
  \bibfield  {author} {\bibinfo {author} {\bibfnamefont {A.}~\bibnamefont
  {Filippetti}}\ and\ \bibinfo {author} {\bibfnamefont {W.~E.}\ \bibnamefont
  {Pickett}},\ }\href {\doibase 10.1103/PhysRevB.62.11571} {\bibfield
  {journal} {\bibinfo  {journal} {Phys. Rev. B}\ }\textbf {\bibinfo {volume}
  {62}},\ \bibinfo {pages} {11571} (\bibinfo {year} {2000})}\BibitemShut
  {NoStop}%
\bibitem [{\citenamefont {Nguyen}\ \emph {et~al.}(2011)\citenamefont {Nguyen},
  \citenamefont {Bach}, \citenamefont {Pham}, \citenamefont {Pham},
  \citenamefont {Nguyen},\ and\ \citenamefont {Hoang}}]{Nguyen11p3613}%
  \BibitemOpen
  \bibfield  {author} {\bibinfo {author} {\bibfnamefont {T.~T.}\ \bibnamefont
  {Nguyen}}, \bibinfo {author} {\bibfnamefont {T.~C.}\ \bibnamefont {Bach}},
  \bibinfo {author} {\bibfnamefont {H.~T.}\ \bibnamefont {Pham}}, \bibinfo
  {author} {\bibfnamefont {T.~T.}\ \bibnamefont {Pham}}, \bibinfo {author}
  {\bibfnamefont {D.~T.}\ \bibnamefont {Nguyen}}, \ and\ \bibinfo {author}
  {\bibfnamefont {N.~N.}\ \bibnamefont {Hoang}},\ }\href@noop {} {\bibfield
  {journal} {\bibinfo  {journal} {Physica B: Condensed Matter}\ }\textbf
  {\bibinfo {volume} {406}},\ \bibinfo {pages} {3613} (\bibinfo {year}
  {2011})}\BibitemShut {NoStop}%
\bibitem [{\citenamefont {Reuter}\ and\ \citenamefont
  {Scheffler}(2001)}]{Reuter01p035406}%
  \BibitemOpen
  \bibfield  {author} {\bibinfo {author} {\bibfnamefont {K.}~\bibnamefont
  {Reuter}}\ and\ \bibinfo {author} {\bibfnamefont {M.}~\bibnamefont
  {Scheffler}},\ }\href@noop {} {\bibfield  {journal} {\bibinfo  {journal}
  {Phys. Rev. B}\ }\textbf {\bibinfo {volume} {65}},\ \bibinfo {pages} {035406}
  (\bibinfo {year} {2001})}\BibitemShut {NoStop}%
\bibitem [{\citenamefont {Perdew}\ \emph {et~al.}(2008)\citenamefont {Perdew},
  \citenamefont {Ruzsinszky}, \citenamefont {G\'abor}, \citenamefont {Vydrov},
  \citenamefont {Scuseria}, \citenamefont {Constantin}, \citenamefont {Zhou},\
  and\ \citenamefont {Burke}}]{Perdew08p136406}%
  \BibitemOpen
  \bibfield  {author} {\bibinfo {author} {\bibfnamefont {J.~P.}~\bibnamefont
  {Perdew}}, \bibinfo {author} {\bibfnamefont {A.}~\bibnamefont {Ruzsinszky}},
  \bibinfo {author} {\bibfnamefont {G.~I.}~\bibnamefont {Csonka}}, \bibinfo
  {author} {\bibfnamefont {O.~A.}~\bibnamefont {Vydrov}}, \bibinfo {author}
  {\bibfnamefont {G.~E.}~\bibnamefont {Scuseria}}, \bibinfo {author}
  {\bibfnamefont {L.~A.}~\bibnamefont {Constantin}}, \bibinfo {author}
  {\bibfnamefont {X.}~\bibnamefont {Zhou}}, \ and\ \bibinfo {author}
  {\bibfnamefont {K.}~\bibnamefont {Burke}},\ }\href@noop {} {\bibfield
  {journal} {\bibinfo  {journal} {Phys. Rev. Lett.}\ }\textbf {\bibinfo
  {volume} {100}},\ \bibinfo {pages} {136406} (\bibinfo {year}
  {2008})}\BibitemShut {NoStop}%
\bibitem [{\citenamefont {Giannozzi}\ \emph {et~al.}(2009)\citenamefont
  {Giannozzi}, \citenamefont {Baroni}, \citenamefont {Bonini}, \citenamefont
  {Calandra}, \citenamefont {Car}, \citenamefont {Cavazzoni}, \citenamefont
  {Ceresoli}, \citenamefont {Chiarotti}, \citenamefont {Cococcioni},
  \citenamefont {Dabo}, \citenamefont {Corso}, \citenamefont {de~Gironcoli},
  \citenamefont {Fabris}, \citenamefont {Fratesi}, \citenamefont {Gebauer},
  \citenamefont {Gerstmann}, \citenamefont {Gougoussis}, \citenamefont
  {Kokalj}, \citenamefont {Lazzeri}, \citenamefont {Martin-Samos},
  \citenamefont {Marzari}, \citenamefont {Mauri}, \citenamefont {Mazzarello},
  \citenamefont {Paolini}, \citenamefont {Pasquarello}, \citenamefont
  {Paulatto}, \citenamefont {Sbraccia}, \citenamefont {Scandolo}, \citenamefont
  {Sclauzero}, \citenamefont {Seitsonen}, \citenamefont {Smogunov},
  \citenamefont {Umari},\ and\ \citenamefont
  {Wentzcovitch}}]{Giannozzi09p395502}%
  \BibitemOpen
  \bibfield  {author} {\bibinfo {author} {\bibfnamefont {P.}~\bibnamefont
  {Giannozzi}}, \bibinfo {author} {\bibfnamefont {S.}~\bibnamefont {Baroni}},
  \bibinfo {author} {\bibfnamefont {N.}~\bibnamefont {Bonini}}, \bibinfo
  {author} {\bibfnamefont {M.}~\bibnamefont {Calandra}}, \bibinfo {author}
  {\bibfnamefont {R.}~\bibnamefont {Car}}, \bibinfo {author} {\bibfnamefont
  {C.}~\bibnamefont {Cavazzoni}}, \bibinfo {author} {\bibfnamefont
  {D.}~\bibnamefont {Ceresoli}}, \bibinfo {author} {\bibfnamefont {G.~L.}\
  \bibnamefont {Chiarotti}}, \bibinfo {author} {\bibfnamefont {M.}~\bibnamefont
  {Cococcioni}}, \bibinfo {author} {\bibfnamefont {I.}~\bibnamefont {Dabo}},
  \bibinfo {author} {\bibfnamefont {A.~D.}\ \bibnamefont {Corso}}, \bibinfo
  {author} {\bibfnamefont {S.}~\bibnamefont {de~Gironcoli}}, \bibinfo {author}
  {\bibfnamefont {S.}~\bibnamefont {Fabris}}, \bibinfo {author} {\bibfnamefont
  {G.}~\bibnamefont {Fratesi}}, \bibinfo {author} {\bibfnamefont
  {R.}~\bibnamefont {Gebauer}}, \bibinfo {author} {\bibfnamefont
  {U.}~\bibnamefont {Gerstmann}}, \bibinfo {author} {\bibfnamefont
  {C.}~\bibnamefont {Gougoussis}}, \bibinfo {author} {\bibfnamefont
  {A.}~\bibnamefont {Kokalj}}, \bibinfo {author} {\bibfnamefont
  {M.}~\bibnamefont {Lazzeri}}, \bibinfo {author} {\bibfnamefont
  {L.}~\bibnamefont {Martin-Samos}}, \bibinfo {author} {\bibfnamefont
  {N.}~\bibnamefont {Marzari}}, \bibinfo {author} {\bibfnamefont
  {F.}~\bibnamefont {Mauri}}, \bibinfo {author} {\bibfnamefont
  {R.}~\bibnamefont {Mazzarello}}, \bibinfo {author} {\bibfnamefont
  {S.}~\bibnamefont {Paolini}}, \bibinfo {author} {\bibfnamefont
  {A.}~\bibnamefont {Pasquarello}}, \bibinfo {author} {\bibfnamefont
  {L.}~\bibnamefont {Paulatto}}, \bibinfo {author} {\bibfnamefont
  {C.}~\bibnamefont {Sbraccia}}, \bibinfo {author} {\bibfnamefont
  {S.}~\bibnamefont {Scandolo}}, \bibinfo {author} {\bibfnamefont
  {G.}~\bibnamefont {Sclauzero}}, \bibinfo {author} {\bibfnamefont {A.~P.}\
  \bibnamefont {Seitsonen}}, \bibinfo {author} {\bibfnamefont {A.}~\bibnamefont
  {Smogunov}}, \bibinfo {author} {\bibfnamefont {P.}~\bibnamefont {Umari}}, \
  and\ \bibinfo {author} {\bibfnamefont {R.~M.}\ \bibnamefont {Wentzcovitch}},\
  }\href@noop {} {\bibfield  {journal} {\bibinfo  {journal} {J. Phys.:Condens.
  Matter}\ }\textbf {\bibinfo {volume} {21}},\ \bibinfo {pages} {395502}
  (\bibinfo {year} {2009})}\BibitemShut {NoStop}%
\bibitem [{\citenamefont {Luo}\ \emph {et~al.}(2007)\citenamefont {Luo},
  \citenamefont {Franceschetti}, \citenamefont {Varela}, \citenamefont {Tao},
  \citenamefont {Pennycook},\ and\ \citenamefont {Pantelides}}]{Luo07p036402}%
  \BibitemOpen
  \bibfield  {author} {\bibinfo {author} {\bibfnamefont {W.}~\bibnamefont
  {Luo}}, \bibinfo {author} {\bibfnamefont {A.}~\bibnamefont {Franceschetti}},
  \bibinfo {author} {\bibfnamefont {M.}~\bibnamefont {Varela}}, \bibinfo
  {author} {\bibfnamefont {J.}~\bibnamefont {Tao}}, \bibinfo {author}
  {\bibfnamefont {S.~J.}\ \bibnamefont {Pennycook}}, \ and\ \bibinfo {author}
  {\bibfnamefont {S.~T.}\ \bibnamefont {Pantelides}},\ }\href {\doibase
  10.1103/PhysRevLett.99.036402} {\bibfield  {journal} {\bibinfo  {journal}
  {Phys. Rev. Lett.}\ }\textbf {\bibinfo {volume} {99}},\ \bibinfo {pages}
  {036402} (\bibinfo {year} {2007})}\BibitemShut {NoStop}%
\bibitem [{\citenamefont {Cococcioni}\ and\ \citenamefont
  {de~Gironcoli}(2005)}]{Cococcioni05p035105}%
  \BibitemOpen
  \bibfield  {author} {\bibinfo {author} {\bibfnamefont {M.}~\bibnamefont
  {Cococcioni}}\ and\ \bibinfo {author} {\bibfnamefont {S.}~\bibnamefont
  {de~Gironcoli}},\ }\href@noop {} {\bibfield  {journal} {\bibinfo  {journal}
  {Phys. Rev. B}\ }\textbf {\bibinfo {volume} {71}},\ \bibinfo {pages} {035105}
  (\bibinfo {year} {2005})}\BibitemShut {NoStop}%
\bibitem [{\citenamefont {Rappe}\ \emph {et~al.}(1990)\citenamefont {Rappe},
  \citenamefont {Rabe}, \citenamefont {Kaxiras},\ and\ \citenamefont
  {Joannopoulos}}]{Rappe90p1227}%
  \BibitemOpen
  \bibfield  {author} {\bibinfo {author} {\bibfnamefont {A.~M.}\ \bibnamefont
  {Rappe}}, \bibinfo {author} {\bibfnamefont {K.~M.}\ \bibnamefont {Rabe}},
  \bibinfo {author} {\bibfnamefont {E.}~\bibnamefont {Kaxiras}}, \ and\
  \bibinfo {author} {\bibfnamefont {J.~D.}\ \bibnamefont {Joannopoulos}},\
  }\href@noop {} {\bibfield  {journal} {\bibinfo  {journal} {Phys. Rev. B Rapid
  Comm.}\ }\textbf {\bibinfo {volume} {41}},\ \bibinfo {pages} {1227} (\bibinfo
  {year} {1990})}\BibitemShut {NoStop}%
\bibitem [{\citenamefont {Ramer}\ and\ \citenamefont
  {Rappe}(1999)}]{Ramer99p12471}%
  \BibitemOpen
  \bibfield  {author} {\bibinfo {author} {\bibfnamefont {N.~J.}\ \bibnamefont
  {Ramer}}\ and\ \bibinfo {author} {\bibfnamefont {A.~M.}\ \bibnamefont
  {Rappe}},\ }\href@noop {} {\bibfield  {journal} {\bibinfo  {journal} {Phys.
  Rev. B}\ }\textbf {\bibinfo {volume} {59}},\ \bibinfo {pages} {12471}
  (\bibinfo {year} {1999})}\BibitemShut {NoStop}%
\bibitem [{Opi()}]{Opium}%
  \BibitemOpen
  \href@noop {} {}\bibinfo {howpublished}
  {http://opium.sourceforge.net}\BibitemShut {NoStop}%
\bibitem [{\citenamefont {Fuchs}\ and\ \citenamefont
  {Scheffler}(1999)}]{Fuchs99p67}%
  \BibitemOpen
  \bibfield  {author} {\bibinfo {author} {\bibfnamefont {M.}~\bibnamefont
  {Fuchs}}\ and\ \bibinfo {author} {\bibfnamefont {M.}~\bibnamefont
  {Scheffler}},\ }\href@noop {} {\bibfield  {journal} {\bibinfo  {journal}
  {Comput. Phys. Commun.}\ }\textbf {\bibinfo {volume} {119}},\ \bibinfo
  {pages} {67} (\bibinfo {year} {1999})}\BibitemShut {NoStop}%
\bibitem [{\citenamefont {Porezag}\ \emph {et~al.}(1999)\citenamefont
  {Porezag}, \citenamefont {Pederson},\ and\ \citenamefont
  {Liu}}]{Porezag99p14132}%
  \BibitemOpen
  \bibfield  {author} {\bibinfo {author} {\bibfnamefont {D.}~\bibnamefont
  {Porezag}}, \bibinfo {author} {\bibfnamefont {M.~R.}\ \bibnamefont
  {Pederson}}, \ and\ \bibinfo {author} {\bibfnamefont {A.~Y.}\ \bibnamefont
  {Liu}},\ }\href@noop {} {\bibfield  {journal} {\bibinfo  {journal} {Phys.
  Rev. B}\ }\textbf {\bibinfo {volume} {60}},\ \bibinfo {pages} {14132}
  (\bibinfo {year} {1999})}\BibitemShut {NoStop}%
\bibitem [{\citenamefont {Poeppelmeier}\ \emph {et~al.}(1982)\citenamefont
  {Poeppelmeier}, \citenamefont {Leonowicz}, \citenamefont {Scanlon},
  \citenamefont {Longo},\ and\ \citenamefont {Longo}}]{Poeppelmeier82p71}%
  \BibitemOpen
  \bibfield  {author} {\bibinfo {author} {\bibfnamefont {K.~R.}\ \bibnamefont
  {Poeppelmeier}}, \bibinfo {author} {\bibfnamefont {M.~E.}\ \bibnamefont
  {Leonowicz}}, \bibinfo {author} {\bibfnamefont {J.~C.}\ \bibnamefont
  {Scanlon}}, \bibinfo {author} {\bibfnamefont {J.~M.}\ \bibnamefont {Longo}},
  \ and\ \bibinfo {author} {\bibfnamefont {W.~B.}\ \bibnamefont {Longo}},\
  }\href@noop {} {\bibfield  {journal} {\bibinfo  {journal} {J. Solid State
  Chem.}\ }\textbf {\bibinfo {volume} {45}},\ \bibinfo {pages} {71} (\bibinfo
  {year} {1982})}\BibitemShut {NoStop}%
\bibitem [{\citenamefont {Bhattacharjee}\ \emph {et~al.}(2008)\citenamefont
  {Bhattacharjee}, \citenamefont {Bousquet},\ and\ \citenamefont
  {Ghosez}}]{Bhattacharjee08p255229}%
  \BibitemOpen
  \bibfield  {author} {\bibinfo {author} {\bibfnamefont {S.}~\bibnamefont
  {Bhattacharjee}}, \bibinfo {author} {\bibfnamefont {E.}~\bibnamefont
  {Bousquet}}, \ and\ \bibinfo {author} {\bibfnamefont {P.}~\bibnamefont
  {Ghosez}},\ }\href@noop {} {\bibfield  {journal} {\bibinfo  {journal} {J.
  Phys.: Condens. Matter}\ }\textbf {\bibinfo {volume} {20}},\ \bibinfo {pages}
  {255229} (\bibinfo {year} {2008})}\BibitemShut {NoStop}%
\bibitem [{\citenamefont {Monkhorst}\ and\ \citenamefont
  {Pack}(1976)}]{Monkhorst76p5188}%
  \BibitemOpen
  \bibfield  {author} {\bibinfo {author} {\bibfnamefont {H.~J.}\ \bibnamefont
  {Monkhorst}}\ and\ \bibinfo {author} {\bibfnamefont {J.~D.}\ \bibnamefont
  {Pack}},\ }\href@noop {} {\bibfield  {journal} {\bibinfo  {journal} {Phys.
  Rev. B}\ }\textbf {\bibinfo {volume} {13}},\ \bibinfo {pages} {5188}
  (\bibinfo {year} {1976})}\BibitemShut {NoStop}%
\bibitem [{\citenamefont {Baroni}\ \emph {et~al.}(2001)\citenamefont {Baroni},
  \citenamefont {de~Gironcoli},\ and\ \citenamefont
  {Dal~Corso}}]{Baroni01p515}%
  \BibitemOpen
  \bibfield  {author} {\bibinfo {author} {\bibfnamefont {S.}~\bibnamefont
  {Baroni}}, \bibinfo {author} {\bibfnamefont {S.}~\bibnamefont
  {de~Gironcoli}}, \ and\ \bibinfo {author} {\bibfnamefont {A.}~\bibnamefont
  {Dal~Corso}},\ }\href@noop {} {\bibfield  {journal} {\bibinfo  {journal}
  {Rev. Mod. Phys.}\ }\textbf {\bibinfo {volume} {73}},\ \bibinfo {pages} {515}
  (\bibinfo {year} {2001})}\BibitemShut {NoStop}%
\bibitem [{\citenamefont {Gonze}(1995)}]{Gonze95p1096}%
  \BibitemOpen
  \bibfield  {author} {\bibinfo {author} {\bibfnamefont {X.}~\bibnamefont
  {Gonze}},\ }\href@noop {} {\bibfield  {journal} {\bibinfo  {journal} {Phys.
  Rev. A}\ }\textbf {\bibinfo {volume} {52}},\ \bibinfo {pages} {1096}
  (\bibinfo {year} {1995})}\BibitemShut {NoStop}%
\bibitem [{\citenamefont {Tasker}(1979)}]{Tasker79p4977}%
  \BibitemOpen
  \bibfield  {author} {\bibinfo {author} {\bibfnamefont {P.~W.}\ \bibnamefont
  {Tasker}},\ }\href {http://stacks.iop.org/0022-3719/12/i=22/a=036} {\bibfield
   {journal} {\bibinfo  {journal} {Journal of Physics C: Solid State Physics}\
  }\textbf {\bibinfo {volume} {12}},\ \bibinfo {pages} {4977} (\bibinfo {year}
  {1979})}\BibitemShut {NoStop}%
\bibitem [{\citenamefont {Pickett}\ and\ \citenamefont
  {Singh}(1996)}]{Pickett96p1146}%
  \BibitemOpen
  \bibfield  {author} {\bibinfo {author} {\bibfnamefont {W.~E.}\ \bibnamefont
  {Pickett}}\ and\ \bibinfo {author} {\bibfnamefont {D.~J.}\ \bibnamefont
  {Singh}},\ }\href {\doibase 10.1103/PhysRevB.53.1146} {\bibfield  {journal}
  {\bibinfo  {journal} {Phys. Rev. B}\ }\textbf {\bibinfo {volume} {53}},\
  \bibinfo {pages} {1146} (\bibinfo {year} {1996})}\BibitemShut {NoStop}%
\bibitem [{\citenamefont {Jung}\ \emph {et~al.}(1997)\citenamefont {Jung},
  \citenamefont {Kim}, \citenamefont {Eom}, \citenamefont {Noh}, \citenamefont
  {Choi}, \citenamefont {Yu}, \citenamefont {Kwon},\ and\ \citenamefont
  {Chung}}]{Jung97p15489}%
  \BibitemOpen
  \bibfield  {author} {\bibinfo {author} {\bibfnamefont {J.~H.}\ \bibnamefont
  {Jung}}, \bibinfo {author} {\bibfnamefont {K.~H.}\ \bibnamefont {Kim}},
  \bibinfo {author} {\bibfnamefont {D.~J.}\ \bibnamefont {Eom}}, \bibinfo
  {author} {\bibfnamefont {T.~W.}\ \bibnamefont {Noh}}, \bibinfo {author}
  {\bibfnamefont {E.~J.}\ \bibnamefont {Choi}}, \bibinfo {author}
  {\bibfnamefont {J.}~\bibnamefont {Yu}}, \bibinfo {author} {\bibfnamefont
  {Y.~S.}\ \bibnamefont {Kwon}}, \ and\ \bibinfo {author} {\bibfnamefont
  {Y.}~\bibnamefont {Chung}},\ }\href {\doibase 10.1103/PhysRevB.55.15489}
  {\bibfield  {journal} {\bibinfo  {journal} {Phys. Rev. B}\ }\textbf {\bibinfo
  {volume} {55}},\ \bibinfo {pages} {15489} (\bibinfo {year}
  {1997})}\BibitemShut {NoStop}%
\bibitem [{\citenamefont {Tsukahara}\ \emph {et~al.}(2010)\citenamefont
  {Tsukahara}, \citenamefont {Ishibashi},\ and\ \citenamefont
  {Terakura}}]{Tsukahara10p214108}%
  \BibitemOpen
  \bibfield  {author} {\bibinfo {author} {\bibfnamefont {H.}~\bibnamefont
  {Tsukahara}}, \bibinfo {author} {\bibfnamefont {S.}~\bibnamefont
  {Ishibashi}}, \ and\ \bibinfo {author} {\bibfnamefont {K.}~\bibnamefont
  {Terakura}},\ }\href {\doibase 10.1103/PhysRevB.81.214108} {\bibfield
  {journal} {\bibinfo  {journal} {Phys. Rev. B}\ }\textbf {\bibinfo {volume}
  {81}},\ \bibinfo {pages} {214108} (\bibinfo {year} {2010})}\BibitemShut
  {NoStop}%
\bibitem [{\citenamefont {Cahn}(1973)}]{Cahn73p3}%
  \BibitemOpen
  \bibfield  {author} {\bibinfo {author} {\bibfnamefont {J.~W.}\ \bibnamefont
  {Cahn}},\ }\href@noop {} {\emph {\bibinfo {title} {Interfacial
  Segregation}}}\ (\bibinfo  {publisher} {ASM, Metals, Park, OH},\ \bibinfo
  {year} {1973})\ p.~\bibinfo {pages} {3}\BibitemShut {NoStop}%
\bibitem [{\citenamefont {Lee}\ \emph {et~al.}(2011)\citenamefont {Lee},
  \citenamefont {Behera}, \citenamefont {Wachsman}, \citenamefont {Phillpot},\
  and\ \citenamefont {Sinnott}}]{Lee11p115418}%
  \BibitemOpen
  \bibfield  {author} {\bibinfo {author} {\bibfnamefont {C.-W.}\ \bibnamefont
  {Lee}}, \bibinfo {author} {\bibfnamefont {R.~K.}\ \bibnamefont {Behera}},
  \bibinfo {author} {\bibfnamefont {E.~D.}\ \bibnamefont {Wachsman}}, \bibinfo
  {author} {\bibfnamefont {S.~R.}\ \bibnamefont {Phillpot}}, \ and\ \bibinfo
  {author} {\bibfnamefont {S.~B.}\ \bibnamefont {Sinnott}},\ }\href@noop {}
  {\bibfield  {journal} {\bibinfo  {journal} {Phys. Rev. B}\ }\textbf {\bibinfo
  {volume} {83}},\ \bibinfo {pages} {115418} (\bibinfo {year}
  {2011})}\BibitemShut {NoStop}%
\bibitem [{\citenamefont {Heifets}\ \emph {et~al.}(2011)\citenamefont
  {Heifets}, \citenamefont {Kotomin}, \citenamefont {Mastrikov}, \citenamefont
  {Piskunov},\ and\ \citenamefont {Maier}}]{Heifets11p491}%
  \BibitemOpen
  \bibfield  {author} {\bibinfo {author} {\bibfnamefont {E.}~\bibnamefont
  {Heifets}}, \bibinfo {author} {\bibfnamefont {E.~A.}\ \bibnamefont
  {Kotomin}}, \bibinfo {author} {\bibfnamefont {Y.~A.}\ \bibnamefont
  {Mastrikov}}, \bibinfo {author} {\bibfnamefont {S.}~\bibnamefont {Piskunov}},
  \ and\ \bibinfo {author} {\bibfnamefont {J.}~\bibnamefont {Maier}},\
  }\href@noop {} {\emph {\bibinfo {title} {Thermodynamics - Interaction studies
  - Solids, liquids and gases}}}\ (\bibinfo  {publisher} {InTech Open Access
  Publishers},\ \bibinfo {year} {2011})\ pp.\ \bibinfo {pages}
  {491--518}\BibitemShut {NoStop}%
\bibitem [{\citenamefont {Kolpak}\ \emph {et~al.}(2008)\citenamefont {Kolpak},
  \citenamefont {Li}, \citenamefont {Shao}, \citenamefont {Rappe},\ and\
  \citenamefont {Bonnell}}]{Kolpak08p036102}%
  \BibitemOpen
  \bibfield  {author} {\bibinfo {author} {\bibfnamefont {A.~M.}\ \bibnamefont
  {Kolpak}}, \bibinfo {author} {\bibfnamefont {D.}~\bibnamefont {Li}}, \bibinfo
  {author} {\bibfnamefont {R.}~\bibnamefont {Shao}}, \bibinfo {author}
  {\bibfnamefont {A.~M.}\ \bibnamefont {Rappe}}, \ and\ \bibinfo {author}
  {\bibfnamefont {D.~A.}\ \bibnamefont {Bonnell}},\ }\href@noop {} {\bibfield
  {journal} {\bibinfo  {journal} {Phys. Rev. Lett.}\ }\textbf {\bibinfo
  {volume} {101}},\ \bibinfo {pages} {036102} (\bibinfo {year}
  {2008})}\BibitemShut {NoStop}%
\bibitem [{\citenamefont {Levchenko}\ and\ \citenamefont
  {Rappe}(2008)}]{Levchenko08p256101}%
  \BibitemOpen
  \bibfield  {author} {\bibinfo {author} {\bibfnamefont {S.~V.}\ \bibnamefont
  {Levchenko}}\ and\ \bibinfo {author} {\bibfnamefont {A.~M.}\ \bibnamefont
  {Rappe}},\ }\href@noop {} {\bibfield  {journal} {\bibinfo  {journal} {Phys.
  Rev. Lett.}\ }\textbf {\bibinfo {volume} {100}},\ \bibinfo {pages} {256101}
  (\bibinfo {year} {2008})}\BibitemShut {NoStop}%
\bibitem [{\citenamefont {Horowitz}\ and\ \citenamefont
  {Longo}(1978)}]{Horowitz78p1359}%
  \BibitemOpen
  \bibfield  {author} {\bibinfo {author} {\bibfnamefont {H.~S.}\ \bibnamefont
  {Horowitz}}\ and\ \bibinfo {author} {\bibfnamefont {J.~M.}\ \bibnamefont
  {Longo}},\ }\href
  {http://www.sciencedirect.com/science/article/pii/0025540878901277}
  {\bibfield  {journal} {\bibinfo  {journal} {Materials Research Bulletin}\
  }\textbf {\bibinfo {volume} {13}},\ \bibinfo {pages} {1359} (\bibinfo {year}
  {1978})}\BibitemShut {NoStop}%
\bibitem [{\citenamefont {Balakirev}\ and\ \citenamefont
  {Golikov}(2006)}]{Balakirev06pS49}%
  \BibitemOpen
  \bibfield  {author} {\bibinfo {author} {\bibfnamefont {V.~F.}\ \bibnamefont
  {Balakirev}}\ and\ \bibinfo {author} {\bibfnamefont {Y.~V.}\ \bibnamefont
  {Golikov}},\ }\href@noop {} {\bibfield  {journal} {\bibinfo  {journal}
  {Inorganic Materials}\ }\textbf {\bibinfo {volume} {42}},\ \bibinfo {pages}
  {S49} (\bibinfo {year} {2006})}\BibitemShut {NoStop}%
\bibitem [{\citenamefont {Furche}(2001)}]{Furche01p195120}%
  \BibitemOpen
  \bibfield  {author} {\bibinfo {author} {\bibfnamefont {F.}~\bibnamefont
  {Furche}},\ }\href@noop {} {\bibfield  {journal} {\bibinfo  {journal} {Phys.
  Rev. B}\ }\textbf {\bibinfo {volume} {64}},\ \bibinfo {pages} {195120}
  (\bibinfo {year} {2001})}\BibitemShut {NoStop}%
\bibitem [{\citenamefont {Walter}\ and\ \citenamefont
  {Rappe}(1999)}]{Walter99p11}%
  \BibitemOpen
  \bibfield  {author} {\bibinfo {author} {\bibfnamefont {E.~J.}\ \bibnamefont
  {Walter}}\ and\ \bibinfo {author} {\bibfnamefont {A.~M.}\ \bibnamefont
  {Rappe}},\ }\href@noop {} {\bibfield  {journal} {\bibinfo  {journal} {Surf.
  Sci.}\ }\textbf {\bibinfo {volume} {427}},\ \bibinfo {pages} {11} (\bibinfo
  {year} {1999})}\BibitemShut {NoStop}%
\bibitem [{\citenamefont {Grinberg}\ \emph {et~al.}(2002)\citenamefont
  {Grinberg}, \citenamefont {Yourdshahyan},\ and\ \citenamefont
  {Rappe}}]{Grinberg02p2264}%
  \BibitemOpen
  \bibfield  {author} {\bibinfo {author} {\bibfnamefont {I.}~\bibnamefont
  {Grinberg}}, \bibinfo {author} {\bibfnamefont {Y.}~\bibnamefont
  {Yourdshahyan}}, \ and\ \bibinfo {author} {\bibfnamefont {A.~M.}\
  \bibnamefont {Rappe}},\ }\href@noop {} {\bibfield  {journal} {\bibinfo
  {journal} {J. Chem. Phys.}\ }\textbf {\bibinfo {volume} {117}},\ \bibinfo
  {pages} {2264} (\bibinfo {year} {2002})}\BibitemShut {NoStop}%
\bibitem [{\citenamefont {Fritsch}\ and\ \citenamefont
  {Navrotsky}(1996)}]{Fritsch96p1761}%
  \BibitemOpen
  \bibfield  {author} {\bibinfo {author} {\bibfnamefont {S.}~\bibnamefont
  {Fritsch}}\ and\ \bibinfo {author} {\bibfnamefont {A.}~\bibnamefont
  {Navrotsky}},\ }\href@noop {} {\bibfield  {journal} {\bibinfo  {journal}
  {Journal of the American Ceramic Society}\ }\textbf {\bibinfo {volume}
  {79}},\ \bibinfo {pages} {1761} (\bibinfo {year} {1996})}\BibitemShut
  {NoStop}%
\bibitem [{\citenamefont {Kotomin}\ \emph {et~al.}(2008)\citenamefont
  {Kotomin}, \citenamefont {Mastrikov}, \citenamefont {Heifets},\ and\
  \citenamefont {Maier}}]{Kotomin08p4644}%
  \BibitemOpen
  \bibfield  {author} {\bibinfo {author} {\bibfnamefont {E.~A.}\ \bibnamefont
  {Kotomin}}, \bibinfo {author} {\bibfnamefont {Y.~A.}\ \bibnamefont
  {Mastrikov}}, \bibinfo {author} {\bibfnamefont {E.}~\bibnamefont {Heifets}},
  \ and\ \bibinfo {author} {\bibfnamefont {J.}~\bibnamefont {Maier}},\ }\href
  {\doibase 10.1039/B804378G} {\bibfield  {journal} {\bibinfo  {journal} {Phys.
  Chem. Chem. Phys.}\ }\textbf {\bibinfo {volume} {10}},\ \bibinfo {pages}
  {4644} (\bibinfo {year} {2008})}\BibitemShut {NoStop}%
\bibitem [{\citenamefont {Jr.}(1998)}]{Chase98p1}%
  \BibitemOpen
  \bibfield  {author} {\bibinfo {author} {\bibfnamefont {M.~W.}\
  \bibnamefont {Chase Jr.}},\ }\href@noop {} {\bibfield  {journal} {\bibinfo
  {journal} {J. Phys. Chem. Ref. Data, Monograph 9}\ ,\ \bibinfo {pages} {1}}
  (\bibinfo {year} {1998})}\BibitemShut {NoStop}%
\bibitem [{\citenamefont {Fawcett}\ \emph {et~al.}(1998)\citenamefont
  {Fawcett}, \citenamefont {Sunstrom~IV}, \citenamefont {Greenblatt},
  \citenamefont {Croft},\ and\ \citenamefont {Ramanujachary}}]{Fawcett98p3643}%
  \BibitemOpen
  \bibfield  {author} {\bibinfo {author} {\bibfnamefont {I.~D.}\ \bibnamefont
  {Fawcett}}, \bibinfo {author} {\bibfnamefont {J.~E.}\ \bibnamefont
  {Sunstrom~IV}}, \bibinfo {author} {\bibfnamefont {M.}~\bibnamefont
  {Greenblatt}}, \bibinfo {author} {\bibfnamefont {M.}~\bibnamefont {Croft}}, \
  and\ \bibinfo {author} {\bibfnamefont {K.~V.}\ \bibnamefont
  {Ramanujachary}},\ }\href@noop {} {\bibfield  {journal} {\bibinfo  {journal}
  {Chemistry of Materials}\ }\textbf {\bibinfo {volume} {10}},\ \bibinfo
  {pages} {3643} (\bibinfo {year} {1998})}\BibitemShut {NoStop}%
\bibitem [{\citenamefont {Ling}\ \emph {et~al.}(2001)\citenamefont {Ling},
  \citenamefont {Neumeier},\ and\ \citenamefont {Argyriou}}]{Ling01p167}%
  \BibitemOpen
  \bibfield  {author} {\bibinfo {author} {\bibfnamefont {C.~D.}\ \bibnamefont
  {Ling}}, \bibinfo {author} {\bibfnamefont {J.}~\bibnamefont {Neumeier}}, \
  and\ \bibinfo {author} {\bibfnamefont {D.~N.}\ \bibnamefont {Argyriou}},\
  }\href@noop {} {\bibfield  {journal} {\bibinfo  {journal} {Journal of Solid
  State Chemistry}\ }\textbf {\bibinfo {volume} {160}},\ \bibinfo {pages} {167}
  (\bibinfo {year} {2001})}\BibitemShut {NoStop}%
\bibitem [{\citenamefont {Xue-Wei}\ \emph {et~al.}(2011)\citenamefont
  {Xue-Wei}, \citenamefont {Hai-Xin}, \citenamefont {Xiao-Jun},\ and\
  \citenamefont {Xing-Gan}}]{Wu11p107101}%
  \BibitemOpen
  \bibfield  {author} {\bibinfo {author} {\bibfnamefont {W.}~\bibnamefont
  {Xue-Wei}}, \bibinfo {author} {\bibfnamefont {Z.}~\bibnamefont {Hai-Xin}},
  \bibinfo {author} {\bibfnamefont {L.}~\bibnamefont {Xiao-Jun}}, \ and\
  \bibinfo {author} {\bibfnamefont {Z.}~\bibnamefont {Xing-Gan}},\ }\href
  {http://stacks.iop.org/0256-307X/28/i=10/a=107101} {\bibfield  {journal}
  {\bibinfo  {journal} {Chinese Physics Letters}\ }\textbf {\bibinfo {volume}
  {28}},\ \bibinfo {pages} {107101} (\bibinfo {year} {2011})}\BibitemShut
  {NoStop}%
\bibitem [{\citenamefont {Bochu}\ \emph {et~al.}(1980)\citenamefont {Bochu},
  \citenamefont {Buevoz}, \citenamefont {Chenavas}, \citenamefont {Collomb},
  \citenamefont {Joubert},\ and\ \citenamefont {Marezio}}]{Bochu80p133}%
  \BibitemOpen
  \bibfield  {author} {\bibinfo {author} {\bibfnamefont {B.}~\bibnamefont
  {Bochu}}, \bibinfo {author} {\bibfnamefont {J.}~\bibnamefont {Buevoz}},
  \bibinfo {author} {\bibfnamefont {J.}~\bibnamefont {Chenavas}}, \bibinfo
  {author} {\bibfnamefont {A.}~\bibnamefont {Collomb}}, \bibinfo {author}
  {\bibfnamefont {J.}~\bibnamefont {Joubert}}, \ and\ \bibinfo {author}
  {\bibfnamefont {M.}~\bibnamefont {Marezio}},\ }\href@noop {} {\bibfield
  {journal} {\bibinfo  {journal} {Solid State Communications}\ }\textbf
  {\bibinfo {volume} {36}},\ \bibinfo {pages} {133} (\bibinfo {year}
  {1980})}\BibitemShut {NoStop}%
\bibitem [{\citenamefont {Jorge}\ \emph {et~al.}(2001)\citenamefont {Jorge},
  \citenamefont {dos Santos},\ and\ \citenamefont {Nunes}}]{Melo01p915}%
  \BibitemOpen
  \bibfield  {author} {\bibinfo {author} {\bibfnamefont {M.~E.}\
  \bibnamefont {Melo Jorge}}, \bibinfo {author} {\bibfnamefont {A.}\ \bibnamefont
  {Correia dos Santos}}, \ and\ \bibinfo {author} {\bibfnamefont {M.~R.}\ \bibnamefont
  {Nunes}},\ }\href@noop {} {\bibfield  {journal} {\bibinfo  {journal}
  {International Journal of Inorganic Materials}\ }\textbf {\bibinfo {volume}
  {3}},\ \bibinfo {pages} {915} (\bibinfo {year} {2001})}\BibitemShut {NoStop}%
\end{thebibliography}%

\end{document}